# Row Sampling for Matrix Algorithms via a Non-Commutative Bernstein Bound

Malik Magdon-Ismail
magdon@cs.rpi.edu
Computer Science Department
Rensselaer Ploytechnic Institute
110 8th Street, Troy, NY 12180, USA

August 7, 2018

**Abstract**

We focus the use of *row sampling* for approximating matrix algorithms. We give applications to matrix multipication; sparse matrix reconstruction; and, $\ell_2$ regression. For a matrix $\mathrm{A} \in \mathbb{R}^{m\times d}$ which represents $m$ points in $d \ll m$ dimensions, all of these tasks can be achieved in $O(md^2)$ via the singular value decomposition (SVD). For appropriate row-sampling probabilities (which typically depend on the norms of the rows of the $m \times d$ left singular matrix of A (the *leverage scores*), we give row-sampling algorithms with linear (up to polylog factors) dependence on the stable rank of A. This result is achieved through the application of non-commutative Bernstein bounds.

We then give, to our knowledge, the first algorithms for computing approximations to the appropriate row-sampling probabilities without going through the SVD of A. Thus, these are the first $o(md^2)$ algorithms for row-sampling based approximations to the matrix algorithms which use leverage scores as the sampling probabilities. The techniques we use to approximate sampling according to the leverage scores uses some powerful recent results in the theory of random projections for embedding, and may be of some independent interest. We confess that one may perform all these matrix tasks more efficiently using these same random projection methods, however the resulting algorithms are in terms of a small number of linear combinations of all the rows. In many applications, the actual rows of A have some physical meaning and so methods based on a small number of the actual rows are of interest.

## 1 Introduction

Matrix algorithms (eg. matrix multiplication, SVD, $\ell_2$ regression) are of widespread use in many application areas: data mining (Azar *et al.*, 2001); recommendations systems (Drineas *et al.*, 2002); information retrieval (Berry *et al.*, 1995; Papadimitriou *et al.*, 2000); web search Kleinberg (1999); Achlioptas *et al.* (2001); clustering (Drineas *et al.*, 2004; McSherry, 2001); mixture modeling (Kannan *et al.*, 2008; Achlioptas and McSherry, 2005); etc. Based on the importance of matrix algorithms, there has been considerable research energy expended on breaking the $O(md^2)$ bound required by exact SVD methods (Golub and Van Loan, 1983).

Starting with a seminal result of Frieze *et al.* (1998), a large number of results using non-uniform sampling to speed up matrix computations have appeared (Achlioptas and McSherry, 2007; Deshpande *et al.*, 2006; Deshpande and Vempala, 2006; Drineas *et al.*, 2006a,b,c,d,e; Rudelson and Vershynin, 2007; Magen and Zouzias, 2010), some of which give relative error guarantees (Deshpande *et al.*, 2006; Deshpande and Vempala, 2006; Drineas *et al.*, 2006d,e; Magen and Zouzias, 2010). So far none of the row-sampling based methods which use non-uniform probabilities generated according to the norms of the rows in the left singular matrix have yielded practically efficient algorithms, although their row sampling complexities (number of rows to be sampled) are impressive.

Even more recently, Sarlos (2006) showed how random projections or "sketches" can be used to perform all these tasks efficiently, obtaining the first $o(md^2)$ algorithms when preserving the identity of the rows themselves are not important. In fact, we will find many of these techniques, together with those in Ailon and Chazelle (2006) essential to our algorithm for generating row samples ultimately leading to $o(md^2)$ algorithms based on row-sampling. From now on, we focus on row-sampling algorithms.

We start with the basic result of matrix multiplication. All other results more or less follow from here. In an independent recent work by Magen and Zouzias (2010) which is developed along the lines of using isoperimetric inequalities (Rudelson and Vershynin, 2007) to obtain matrix Chernoff bounds, Magen and Zouzias (2010) show that by sampling nearly a linear number of rows, it is possible to obtain a relative error approximation to matrix multiplication. Specifically, let $\mathrm{A} \in \mathbb{R}^{m\times d_1}$ and $\mathrm{B} \in \mathbb{R}^{m\times d_2}$. Then for $r \geq (\rho \log \frac{d_1+d_2}{\delta})/\epsilon^2$ (where $\rho$ bounds the stable (or "soft") rank of A, B – see later), there is a probability distribution over $\mathcal{I} = \{1, \ldots, m\}$ such that by sampling $r$ rows i.i.d. from $\mathcal{I}$, one can construct sketches $\tilde{\mathrm{A}}$, $\tilde{\mathrm{B}}$ such that $\tilde{\mathrm{A}}^{\mathrm{T}}\tilde{\mathrm{B}} \approx \mathrm{A}^{\mathrm{T}}\mathrm{B}$. Specifically, with probability at least $1-\delta$,

$$\| \tilde{\mathrm{A}}^{\mathrm{T}}\tilde{\mathrm{B}} - \mathrm{A}^{\mathrm{T}}\mathrm{B} \|_2 \leq \epsilon \| \mathrm{A} \|_2 \| \mathrm{B} \|_2.$$

The sampling distribution is relatively simple, relying only on the product of the norms of the rows in A and B. This result is applied to low rank matrix reconstruction and $\ell_2$-regression where the required sampling distribution needs knowledge of the SVD of A and B. It is not known how to sample efficiently from these so-called "expensive" probabilities, because to compute the probabilities requires going through the SVD which inflates the running times.

Our basic result for matrix multiplication is very similar to this, and we arrive at it through a different path using a non-commutative Bernstein bound. Our sampling probabilities are different. In appication of our results to sparse matrix reconstruction and $\ell_2$-regression, the rows of the left singular matrix make an appearance. We show how to approximate these probabilities using random projections at the expense of a poly-logarithmic factor in running times. Specifically, if $\mathbf{u}_1, \ldots, \mathbf{u}_m$ are the rows of the left singular vectors of a matrix A, we show how, in $o(md^2)$, to approximate the "leverage scores" $\mathbf{u}_t^2/d$ to within a poly-log factor. It turns out that such an approximation is sufficient to obtain all the algorithms with only a poly-log bloat in efficiency. Any improvement in the approximation of these probabilities directly translates to improvements in such row-sampling based algorithms. To our knowledge, this is the first attempt to bring row-sampling based algorithms to within the realm of $o(md^2)$. To do so, we will use some powerful results in fast metric preserving embeddings (Johnson and Lindenstrauss (1984); Ailon and Chazelle (2006)).

## 1.1 Basic Notation

Before we can state the results in concrete form, we need some preliminary conventions. In general, $\epsilon \in (0, 1)$ will be an error tolerance parameter; $\beta \in (0, 1]$ is a parameter used to scale probabilities; and, $c, c' > 0$ are generic constants whose value may vary even within different lines of the same derivation. Let $\mathbf{e}_1, \ldots, \mathbf{e}_m$ be the standard basis vectors in $\mathbb{R}^m$. Let $\mathrm{A} \in \mathbb{R}^{m\times d}$ denote an arbitrary matrix which represents $m$ points in $\mathbb{R}^d$. In general, we might represent a matrix such as A (roman, uppercase) by a set of vectors $\mathbf{a}_1, \ldots, \mathbf{a}_m \in \mathbb{R}^d$ (bold, lowercase), so that $\mathrm{A}^{\mathrm{T}} = [\mathbf{a}_1\ \mathbf{a}_2\ \ldots\ \mathbf{a}_m]$; similarly, for a vector $\mathbf{y}$, $\mathbf{y}^{\mathrm{T}} = [y_1, \ldots, y_m]$. Note that $\mathbf{a}_t$ is the $t^{\mathrm{th}}$ row of A, which we may also refer to by $A_{(t)}$; similarly, we may refer to the $t^{\text{th}}$ column as $A^{(t)}$. Let $\text{rank}(A) \leq \min\{m, d\}$ be the rank of A; typically $m \gg d$ and for concreteness, we will assume that $\text{rank}(A) = d$ (all the results easily generalize to $\text{rank}(A) < d$). For matrices, we will use the spectral norm, $\| \cdot \|$; On occasion, we will use the Frobenius norm, $\| \cdot \|_F$. For vectors, $\| \cdot \|_F = \| \cdot \|$ (the standard Euclidean norm). The stable, or "soft" rank, $\rho(A) = \| A \|_F^2 / \| A \|^2 \leq \text{rank}(A)$.

The singular value decomposition (SVD) of A is

$$A = U_A S_A V_A^T.$$

where $U_A$ is an $m \times d$ set of columns which are an orthonotmal basis for the column space in A; $S_A$ is a $d \times d$ positive diagonal matrix of singular values, and V is a $d \times d$ orthogonal matrix. We refer to the singular values of A (the diagonal entries in $S_A$) by $\sigma_i(A)$. We will call a matrix with orthonormal columns an orthonormal matrix; an orthogonal matrix is a square orthonormal matrix. In particular, $U_A^T U_A = V_A^T V_A = V_A V_A^T = I_{d \times d}$. It is possible to extend $U_A$ to a full orthonormal basis of $\mathbb{R}^m$, $[U_A, U_A^\perp]$.

The SVD is important for a number of reasons. The projection of the columns of $A$ onto the $k$ left singular vectors with top $k$ singular values gives the best rank-$k$ approximation to A in the spectral and Frobenius norms. The solution to the linear regression problem is also intimately related to the SVD. In particular, consider the following minimization problem which is minimized at $\mathbf{w}^*$:

$$Z^* = \min_{\mathbf{w}} \| A\mathbf{w} - \mathbf{y} \|^2.$$

**Lemma 1** (Golub and van Loan (1996))**.** *$Z^* = \| U_A^\perp (U_A^\perp)^T \mathbf{y} \|^2$, and $\mathbf{w}^* = V_A S_A^{-1} U_A^T \mathbf{y}$.*

**Row-Sampling Matrices** Our focus is algorithms based on row-sampling. A *row-sampling matrix* $Q \in \mathbb{R}^{r \times m}$ samples $r$ rows of A to form $\tilde{A} = QA$:

$$Q = \begin{bmatrix} \mathbf{r}_1^T \\ \vdots \\ \mathbf{r}_r^T \end{bmatrix}, \qquad \tilde{A} = QA = \begin{bmatrix} \mathbf{r}_1^T A \\ \vdots \\ \mathbf{r}_r^T A \end{bmatrix} = \begin{bmatrix} \lambda_{t_1} \mathbf{a}_{t_1}^T \\ \vdots \\ \lambda_{t_r} \mathbf{a}_{t_r}^T \end{bmatrix},$$

where $\mathbf{r}_j = \lambda_{t_j} \mathbf{e}_{t_j}$; it is easy to verify that the row $\mathbf{r}_j^T A$ samples the $t_j^{\text{th}}$ row of A and rescales it. We are interested in random sampling matrices where each $\mathbf{r}_j$ is i.i.d. according to some distribution. Define a set of sampling probabilities $p_1, \ldots, p_m$, with $p_i > 0$ and $\sum_{i=1}^m p_i = 1$; then $\mathbf{r}_j = \mathbf{e}_t / \sqrt{r p_t}$ with probability $p_t$. Note that the scaling is also related to the sampling probabilities in all the algorithms we consider. We can write $Q^T Q$ as the sum of $r$ independently sampled matrices,

$$Q^T Q = \frac{1}{r} \sum_{j=1}^{r} \mathbf{r}_j \mathbf{r}_j^T$$

where $\mathbf{r}_j \mathbf{r}_j^T$ is a diagonal matrix with only one non-zero diagonal entry; the $t^{th}$ diagonal entry is equal to $1/p_t$ with probability $p_t$. Thus, by construction, for any set of non-zero sampling probabilities, $\mathbb{E}[\mathbf{r}_j \mathbf{r}_j^T] = I_{m \times m}$. Since we are averaging $r$ independent copies, it is reasonable to expect a concentration around the mean, with respect to $r$, and so in some sense, $Q^T Q$ essentially behaves like the identity.

## 1.2 Statement of Results

All the results essentially follow from the following two basic lemmas on how orthonormal subspaces behave with respect to the row-sampling. These are discussed more thoroughly in Section 3, but we state them here sumararily.

**Theorem 2** (Symmetric Orthonormal Subspace Sampling)**.** *Let* $\mathrm{U} \in \mathbb{R}^{m\times d}$ *be orthonormal, and* $\mathrm{S} \in \mathbb{R}^{d\times d}$ *be positive diagonal. Assume the row-sampling probabilities* $p_t$ *satisfy*

$$p_t \geq \beta \frac{\mathbf{u}_t^{\mathrm{T}}\mathrm{S}^2\mathbf{u}_t}{\mathrm{trace}(\mathrm{S}^2)}.$$

*Then, if* $r \geq (4\rho(\mathrm{S})/\beta\epsilon^2)\ln\frac{2d}{\delta}$*, with probability at least* $1-\delta$*,*

$$\| \mathrm{S}^2 - \mathrm{S}\mathrm{U}^{\mathrm{T}}\mathrm{Q}^{\mathrm{T}}\mathrm{Q}\mathrm{U}\mathrm{S} \| \leq \epsilon \| \mathrm{S} \|^2$$

We also have an asymmetric version of this result, which is actually obtained through an application of this result to a composite matrix.

**Theorem 3** (Asymmetric Orthonormal Subspace Sampling)**.** *Let* $\mathrm{W} \in \mathbb{R}^{m\times d_1}$*,* $\mathrm{V} \in \mathbb{R}^{m\times d_2}$ *be orthonormal, and let* $\mathrm{S}_1 \in \mathbb{R}^{d_1\times d_1}$ *and* $\mathrm{S}_2 \in \mathbb{R}^{d_2\times d_2}$ *be two positive diagonal matrices; let* $\rho_i = \rho(\mathrm{S}_i)$*. Consider row sampling probabilities*

$$p_t \geq \beta \frac{\frac{1}{\|\mathrm{S}_1\|^2}\mathbf{w}_t^{\mathrm{T}}\mathrm{S}_1^2\mathbf{w}_t + \frac{1}{\|\mathrm{S}_2\|^2}\mathbf{v}_t^{\mathrm{T}}\mathrm{S}_2^2\mathbf{v}_t}{\rho_1+\rho_2}.$$

*If* $r \geq (8(\rho_1+\rho_2)/\beta\epsilon^2)\ln\frac{2(d_1+d_2)}{\delta}$*, then with probability at least* $1-\delta$*,*

$$\| \mathrm{S}_1\mathrm{W}^{\mathrm{T}}\mathrm{V}\mathrm{S}_2 - \mathrm{S}_1\mathrm{W}^{\mathrm{T}}\mathrm{Q}^{\mathrm{T}}\mathrm{Q}\mathrm{V}\mathrm{S}_2 \| \leq \epsilon \| \mathrm{S}_1 \| \| \mathrm{S}_2 \|$$

We note that these row sampling probabilities are not the usual product row sampling probabilities one uses for matrix multiplication as in Drineas *et al.* (2006a). Unfortunately, there is one small problem with these probabilities. As can be seen, they require some knowledge regarding the spectral norms of $\mathrm{S}_i$. In the statement of these results, since the $\mathrm{S}_i$ are given diagonal matrices, it is easy to compute $\| \mathrm{S}_i \|$. In the application of these results to matrix multiplication, the spectral norm of the input matrices will appear. We will show how to handle this issue later. We now give some applications of these orthonormal subspace sampling results.

**Theorem 4** (Matrix Multiplication in Spectral Norm)**.** *Let* $\mathrm{A} \in \mathbb{R}^{m\times d_1}$ *and* $\mathrm{B} \in \mathbb{R}^{m\times d_2}$ *have rescaled rows* $\hat{\mathbf{a}}_t = \mathbf{a}_t/\| \mathrm{A} \|$ *and* $\hat{\mathbf{b}}_t = \mathbf{b}_t \| \mathrm{B} \|$ *respectively. Let* $\rho_A$ *(resp.* $\rho_B$*) be the stable rank of* A *(resp* B*). Obtain a sampling matrix* $\mathrm{Q} \in \mathbb{R}^{r\times m}$ *using row-sampling probabilities* $p_t$ *satisfying*

$$p_t \geq \beta \frac{\hat{\mathbf{a}}_t^{\mathrm{T}}\hat{\mathbf{a}}_t + \hat{\mathbf{b}}_t^{\mathrm{T}}\hat{\mathbf{b}}_t}{\sum_{t=1}^m \hat{\mathbf{a}}_t^{\mathrm{T}}\hat{\mathbf{a}}_t + \hat{\mathbf{b}}_t^{\mathrm{T}}\hat{\mathbf{b}}_t} = \beta \frac{\hat{\mathbf{a}}_t^{\mathrm{T}}\hat{\mathbf{a}}_t + \hat{\mathbf{b}}_t^{\mathrm{T}}\hat{\mathbf{b}}_t}{\rho_{\mathrm{A}}+\rho_{\mathrm{B}}}.$$

*Then, if* $r \geq \frac{8(\rho_{\mathrm{A}}+\rho_{\mathrm{B}})}{\beta\epsilon^2}\ln\frac{2(d_1+d_2)}{\delta}$*, with probability at least* $1-\delta$*,*

$$\| \mathrm{A}^{\mathrm{T}}\mathrm{B} - \tilde{\mathrm{A}}^{\mathrm{T}}\tilde{\mathrm{B}} \| \leq \epsilon \| \mathrm{A} \| \| \mathrm{B} \|.$$

**Theorem 5** (Sparse Row-Based Matrix Reconstruction)**.** *Let* $\mathrm{A}$ *have the SVD representation* $\mathrm{A} = \mathrm{USV}^{\mathrm{T}}$*, and consider row-sampling probabilities* $p_t$ *satisfying* $p_t \geq \frac{\beta}{d}\mathbf{u}_t^{\mathrm{T}}\mathbf{u}_t$*. Then, if* $r \geq (4(d-\beta)/\beta\epsilon^2)\ln\frac{2d}{\delta}$*, with probability at least* $1-\delta$*,*

$$\|\,\mathrm{A} - \mathrm{A}\tilde{\Pi}_k\,\| \leq \left(\frac{1+\epsilon}{1-\epsilon}\right)^{1/2} \|\,\mathrm{A} - \mathrm{A}_k\,\|,$$

*for* $k = 1, \ldots, d$*, where* $\tilde{\Pi}_k$ *projects onto the top* $k$ *right singular vectors of* $\tilde{\mathrm{A}}$*.*

**Theorem 6** (Relative Error $\ell_2$ Regression)**.** *Let* $\mathrm{A} \in \mathbb{R}^{m\times d}$ *have the SVD representation* $\mathrm{A} = \mathrm{USV}^{\mathrm{T}}$*, and let* $\mathbf{y} \in \mathbb{R}^m$*. Let* $\mathbf{x}^* = \mathrm{A}^+\mathbf{y}$ *be the optimal regression with residual* $\boldsymbol{\epsilon} = \mathbf{y} - \mathrm{A}\mathbf{x}^* = \mathbf{y} - \mathrm{AA}^+\mathbf{y}$*. Assume the sampling probabilities* $p_t$ *satisfy*

$$p_t \geq \beta\left(\frac{\mathbf{u}_t^2}{d} + \frac{(\mathbf{u}_t^2 + \frac{\epsilon_t^2}{\boldsymbol{\epsilon}^{\mathrm{T}}\boldsymbol{\epsilon}})}{d+1} + \frac{\epsilon_t^2}{\boldsymbol{\epsilon}^{\mathrm{T}}\boldsymbol{\epsilon}}\right)$$

*For* $r \geq (8(d+1)/\beta\epsilon^2)\ln\frac{2(d+1)}{\delta}$*, let* $\hat{\mathbf{x}} = (\mathrm{QA})^+\mathrm{Q}\mathbf{y}$ *be the approximate regression. Then, with probability at least* $1-3\delta$*,*

$$\|\,\mathrm{A}\hat{\mathbf{x}} - \mathbf{y}\,\| \leq \left(1 + \epsilon + \epsilon\sqrt{\frac{1+\epsilon}{1-\epsilon}}\right)\|\,\mathrm{A}\mathbf{x}^* - \mathbf{y}\,\|.$$

In addition to sampling according to $\mathbf{u}_t^2$ we also need the residual vector $\boldsymbol{\epsilon} = \mathbf{y} - \mathrm{AA}^+\mathbf{y}$. Unfortunately, we have not yet found an efficient way to get a good approximation (in some form of relative error) to this residual vector.

### 1.2.1 Approximating the Sampling Probabilities

We show how to construct approximate sampling probabilities efficiently so that the algorithms may run in sub-SVD time. The details are given in Section 7.

**Matrix Multiplication** The sampling probabilities depend on $\|\,\mathrm{A}\,\|^2$ and $\|\,\mathrm{B}\,\|^2$. It is possible to get a constant factor approximation to $\|\,\mathrm{A}\,\|^2$ (and similarly $\|\,\mathrm{B}\,\|^2$) with high probability. First sample $\tilde{\mathrm{A}} = \mathrm{QA}$ according to probabilities $p_t = \mathbf{a}_t^2/\|\,\mathrm{A}\,\|_F^2$. These probabilities are easy to compute in $O(md_1)$. By an application of the symmetric subspace sampling theorem (Theorem 20), if $r \geq (4\rho_A/\epsilon^2)\ln\frac{2d_1}{\delta}$, then with probability at least $1-\delta$,

$$(1-\epsilon)\|\,\mathrm{A}\,\|^2 \leq \|\,\tilde{\mathrm{A}}^{\mathrm{T}}\tilde{\mathrm{A}}\,\| \leq (1+\epsilon)\|\,\mathrm{A}\,\|^2.$$

We now run $\Omega(\ln\frac{d_1}{\delta})$ power iterations starting from a random isotropic vector to estimate the spectral norm of $\tilde{\mathrm{A}}^{\mathrm{T}}\tilde{\mathrm{A}}$. The efficiency is $O(md_1 + \rho_A d_1 \ln^2\frac{d_1}{\delta})$.

**Theorem 7.** *With* $r \geq (4\rho_A/\epsilon^2)\ln\frac{2d_1}{\delta}$*, the spectral norm estimate* $\tilde{\sigma}_1^2$ *obtained after* $c\ln\frac{d_1}{\delta}$ *power iterations on* $\tilde{\mathrm{A}}^{\mathrm{T}}\tilde{\mathrm{A}}$ *starting from an isotropic random vector satisfies*

$$\frac{1}{2\sqrt{5}}\|\,\mathrm{A}\,\|^2 \leq \tilde{\sigma}_1^2 \leq \frac{3}{2}\|\,\mathrm{A}\,\|^2.$$

*Further, the estimate* $\tilde{\sigma}_1^2$ *can be computed in* $O(md_1 + \rho_A d_1 \ln^2\frac{d_1}{\delta})$*.*

Note that we only need a constant factor approximation to the spectral norm to get a constant factor approximation to the probabilities, which is all we need for the matrix multiplication algorithm to maintain the same asymptotic efficiency.

**Sparse Row-Based Matrix Reconstruction** We use random embeddings via Fast Johnson-Lindenstrauss Transformations to approximate the probabilities $p_t = \mathbf{u}_t^2/d$. In particular, we show that it is possible to approximately sample in $o(md^2)$ according to the leverage scores. While the efficiency savings are not heroic, this is (to our knowledge) the first sub-SVD algorithm to sample according to the leverage scores.

**Theorem 8.** *There is an algorithm which constructs estimates $p_t$ such that with probability at least $1-\delta$*

$$p_t \geq \frac{1}{c\log^2 m}\frac{\mathbf{u}_t^2}{d}$$

*for $t = 1, \ldots, m$; the running time is in $O\left(md\log m + (md\log d + \frac{md^2}{\log m})\log\frac{1}{\delta}\right)$.*

We note that we have a polylog factor approximation to the probabilities; this results in a polylog bloat to the efficiencies of all the matrix algorithms. Nevertheless, for $m = o(e^d)$, the algorithms remain sub-SVD.

### 1.3 Paper Outline

Next we describe some probabistic tail inequalities which will be useful. We continue with the sampling lemmas for orthonormal matrices, followed by the applications to matrix multiplication, matrix reconstruction and $\ell_2$-regression. Finally, we discuss the algorithms for approximating the sampling probabilities efficiently.

## 2 Probabilistic Tail Inequalities

Since all our arguments involve high probability results, our main bounding tools will be probability tail inequalities. First, let $X_1, \ldots, X_n$ be independent random variables with $\mathbb{E}[X_i] = 0$ and $|\mathbf{X}_i| \leq \gamma$; let $Z_n = \frac{1}{n}\sum_{i=1}^n X_i$. Chernoff, and later Hoeffding gave the bound

**Theorem 9** (Chernoff (1952); Hoeffding (1963))**.** $\mathbb{P}[|Z_n| > \epsilon] \leq 2e^{-n\epsilon^2/2\gamma^2}$.

If in addition one can bound the variance, $\mathbb{E}[X_i^2] \leq s^2$, then we have Bernstein's bound:

**Theorem 10** (Bernstein (1924))**.** $\mathbb{P}[|Z_n| \geq \epsilon] \leq 2e^{-n\epsilon^2/(2s^2+2\gamma\epsilon/3)}$.

Note that when $\epsilon \leq 3s^2/\gamma$, we can simplify the Bernstein bound to $\mathbb{P}[|Z_n| \geq \epsilon] \leq 2e^{-n\epsilon^2/4s^2}$, which is considerably simpler and only involves the variance. The non-commutative versions of these bounds, which extend these inequalities to matrix valued random variables can also be deduced. Let $\mathrm{X}_1, \ldots, \mathrm{X}_n$ be independent copies of a symmetric random matrix $\mathrm{X}$, with $\mathbb{E}[\mathrm{X}] = \mathbf{0}$, and suppose that $\|\mathrm{X}\|_2 \leq \gamma$; let $\mathrm{Z}_n = \frac{1}{n}\sum_{i=1}^n \mathrm{X}_i$. Ahlswede and Winter (2002) gave the fundamental extension of the exponentiation trick for computing Chernoff bounds of scalar random variables to matrix valued random variables (for a simplified proof, see Wigderson and Xiao (2008)):

$$\mathbb{P}\left[\|\mathrm{Z}_n\|_2 > \epsilon\right] \leq \inf_t 2de^{-n\epsilon t/\gamma}\|\ \mathbb{E}\left[e^{t\mathrm{X}/\gamma}\right]\|_2^n. \tag{1}$$

By standard optimization of this bound, one readily obtains the non-commutative tail inequality

**Theorem 11** (Ahlswede and Winter (2002))**.** $\mathbb{P}[\|\mathrm{Z}_n\|_2 > \epsilon] \leq 2de^{-n\epsilon^2/4\gamma^2}$.

*Proof.* The statement is trivial if $\epsilon \geq \gamma$, so assume $\epsilon < \gamma$. The lemma follows from (1) and the following sequence after setting $t = \epsilon/2\gamma \leq \frac{1}{2}$:

$$\| \mathbb{E}\,[e^{t\mathrm{X}/\gamma}]\,\|_2 \overset{(a)}{\leq} 1 + \sum_{\ell=2}^{\infty} \frac{t^\ell}{\ell!}\,\mathbb{E}\,[\|\,(\mathrm{X}/\gamma)^\ell\,\|_2] \overset{(b)}{\leq} 1 + t^2 \leq e^{t^2}, \tag{2}$$

where (a) follows from $\mathbb{E}[\mathrm{X}] = 0$, the triangle inequality and $\|\,\mathbb{E}\,[\cdot]\,\|_2 \leq \mathbb{E}[\|\,\cdot\,\|_2]$; (b) follows because $\|\,\mathrm{X}/\gamma\,\|_2 \leq 1$ and $t \leq \frac{1}{2}$. ■

(We have stated a simplified version of the bound, without taking care to optimize the constants.) As mentioned in Gross *et al.* (2009), one can obtain a non-commuting version of Bernstein's inequality in a similar fashion using (1). Assume that $\|\,\mathbb{E}\,\mathrm{X}^{\mathrm{T}}\mathrm{X}\,\|_2 \leq s^2$. By adapting the standard Bernstein bounding argument to matrices, we have

**Lemma 12.** $\|\,\mathbb{E}\,[e^{t\mathrm{X}/\gamma}]\,\|_2 \leq \exp\left(\frac{s^2}{\gamma^2}(e^t - 1 - t)\right)$.

*Proof.* As in (2), but using (via submultiplicativity) $\|\,(\mathrm{X}/\gamma)^\ell\,\|_2 \leq s^2\gamma^{\ell-2}/\gamma^\ell = s^2/\gamma^2$,

$$\|\,\mathbb{E}\,[e^{t\mathrm{X}/\gamma}]\,\|_2 \leq 1 + \frac{s^2}{\gamma^2}\sum_{\ell=2}^{\infty}\frac{t^\ell}{\ell!} = 1 + \frac{s^2}{\gamma^2}(e^t - 1 - t) \leq \exp\left(\frac{s^2}{\gamma^2}(e^t - 1 - t)\right).$$

■

Using Lemma 12 in (1) with $t = \ln(1 + \epsilon\gamma/s^2)$, and using $(1 + x)\ln(1 + \frac{1}{x}) - 1 \geq \frac{1}{2x+2/3}$, we obtain the following result.

**Theorem 13** (Non-commutative Bernstein)**.** $\mathbb{P}[\|\,\mathrm{Z}_n\,\|_2 > \epsilon] \leq 2de^{-n\epsilon^2/(2s^2+2\gamma\epsilon/3)}$.

Gross *et al.* (2009) gives a simpler version of the non-commutative Bernstein inequality. If $\mathrm{X} \in \mathbb{R}^{d_1 \times d_2}$ is not symmetric, then by considering

$$\begin{bmatrix} \mathbf{0}_{d_1\times d_1} & \mathrm{X} \\ \mathrm{X}^{\mathrm{T}} & \mathbf{0}_{d_2\times d_2} \end{bmatrix},$$

one can get a non-symmetric verision of the non-commutative Chernoff and Bernstein bounds,

**Theorem 14** (Recht (2009))**.** $\mathbb{P}[\|\,\mathrm{Z}_n\,\|_2 > \epsilon] \leq (d_1 + d_2)e^{-n\epsilon^2/(2s^2+2\gamma\epsilon/3)}$.

For most of our purposes, we will only need the symmetric version; again, if $\epsilon \leq 3s^2/\gamma$, then we have the much simpler bound $\mathbb{P}[\|\,\mathrm{Z}_n\,\|_2 > \epsilon] \leq 2de^{-n\epsilon^2/4s^2}$.

# 3 Orthonormal Sampling Lemmas

Let $\mathrm{U} \in \mathbb{R}^{m\times d}$ be an orthonormal matrix, and let $\mathrm{S} \in \mathbb{R}^{d\times d}$ be a diagonal matrix. We are interested in the product $\mathrm{US} \in \mathbb{R}^{m\times d}$; US is the matrix with columns $\mathrm{U}^{(i)}S_{ii}$. Without loss of generality, we can assume that S is positive by flipping the signs of the appropriate columns of U. The row-representation of U is $\mathrm{U}^{\mathrm{T}} = [\mathbf{u}_1, \ldots, \mathbf{u}_m]$; we consider the row sampling probabilities

$$p_t \geq \beta\frac{\mathbf{u}_t^{\mathrm{T}}\mathrm{S}^2\mathbf{u}_t}{\mathrm{trace}(\mathrm{S}^2)}.$$

Since $\mathrm{U}^{\mathrm{T}}\mathrm{U} = \mathrm{I}_{d\times d}$, one can verify that $\mathrm{trace}(\mathrm{S}^2) = \sum_t \mathbf{u}_t^{\mathrm{T}}\mathrm{S}^2\mathbf{u}_t$ is the correct normalization.

**Lemma 15** (Symmetric Subspace Sampling Lemma)**.**

$$\begin{aligned}\mathbb{P}[\|\mathrm{S}^2 - \mathrm{S}\mathrm{U}^{\mathrm{T}}\mathrm{Q}^{\mathrm{T}}\mathrm{Q}\mathrm{U}\mathrm{S}\| > \epsilon\|\mathrm{S}\|^2] &\leq 2d\cdot\exp\left(\frac{-r\epsilon^2}{2(\rho/\beta - \kappa^{-4} + \epsilon(\rho/\beta - \kappa^{-2})/3)}\right),\\ &\leq 2d\cdot\exp\left(\frac{-r\beta\epsilon^2}{4\rho}\right),\end{aligned}$$

*where $\rho$ is the numerical (stable) rank of* $\mathrm{S}$*,* $\rho(\mathrm{S}) = \|\mathrm{S}\|_F^2/\|\mathrm{S}\|^2$*, and $\kappa$ is the condition number,* $\kappa(\mathrm{S}) = \sigma_{\max}(\mathrm{S})/\sigma_{\min}(\mathrm{S})$*.*

**Remarks.** The stable rank $\rho \leq d$ measures the effective dimension of the matrix. The condition number $\kappa \geq 1$, hence the simpler version of the bound, which is valid for $\epsilon \leq 3$. It immediately follows that if $r \geq (4\rho/\beta\epsilon^2)\ln\frac{2d}{\delta}$, then with probability at least $1-\delta$,

$$\|\mathrm{S}^2 - \mathrm{S}\mathrm{U}^{\mathrm{T}}\mathrm{Q}^{\mathrm{T}}\mathrm{Q}\mathrm{U}\mathrm{S}\| \leq \epsilon\|\mathrm{S}\|^2$$

An important special case is when $\mathrm{S} = \mathrm{I}_{d\times d}$, in which case $\rho = d$, $\kappa = 1$ and $\|\mathrm{S}\| = 1$.

**Corollary 16.** *For sampling probabilities* $p_t \geq \frac{\beta}{d}\mathbf{u}_t^{\mathrm{T}}\mathbf{u}_t$*,*

$$\mathbb{P}[\|\mathrm{I} - \mathrm{U}^{\mathrm{T}}\mathrm{Q}^{\mathrm{T}}\mathrm{Q}\mathrm{U}\| > \epsilon] \leq 2d\cdot\exp\left(\frac{-\beta r\epsilon^2}{4(d-\beta)}\right),$$

*Proof.* (of Lemma 15) Note that $\mathrm{U}^{\mathrm{T}}\mathrm{Q}^{\mathrm{T}}\mathrm{Q}\mathrm{U} = \frac{1}{r}\sum_{i=1}^{r}\mathbf{u}_{t_i}\mathbf{u}_{t_i}^{\mathrm{T}}/p_{t_i}$, where $t_i \in [1,m]$ is chosen according to the probability $p_{t_i}$. It follows that

$$\mathrm{S}^2 - \mathrm{S}\mathrm{U}^{\mathrm{T}}\mathrm{Q}^{\mathrm{T}}\mathrm{Q}\mathrm{U}\mathrm{S} = \frac{1}{r}\sum_{i=1}^{r}\mathrm{S}^2 - \frac{1}{p_{t_i}}\mathrm{S}\mathbf{u}_{t_i}\mathbf{u}_{t_i}^{\mathrm{T}}\mathrm{S} = \frac{1}{r}\sum_{i=1}\mathrm{X}_i,$$

where $\mathrm{X}_i$ are independent copies of a matrix-random variable $\mathrm{X} \sim \mathrm{S}^2 - \mathrm{S}\mathbf{u}\mathbf{u}^{\mathrm{T}}\mathrm{S}^2/p$. We prove the following three claims:

$$\begin{aligned}&(i)\ \ \mathbb{E}[\mathrm{X}] = \mathbf{0};\\ &(ii)\ \|\mathrm{X}\| \leq \|\mathrm{S}\|^2(\rho/\beta - \kappa^{-2});\\ &(ii)\ \|\,\mathbb{E}\,\mathrm{X}^{\mathrm{T}}\mathrm{X}\| \leq \|\mathrm{S}\|^4(\rho/\beta - \kappa^{-4}).\end{aligned}$$

The Lemma follows from the non-commutative Bernstein bound with $\epsilon$ replaced by $\epsilon\|\mathrm{S}\|^2$. To prove $(i)$, note that $\mathbb{E}[\mathrm{X}] = \mathrm{S}^2 - \mathrm{S}\,\mathbb{E}\,[\mathbf{u}\mathbf{u}^{\mathrm{T}}/p]\mathrm{S} = \mathrm{S}^2 - \mathrm{S}\left(\sum_{t=1}^{m}\mathbf{u}_t\mathbf{u}_t^{\mathrm{T}}\right)\mathrm{S} = \mathbf{0}$, because $\sum_{t=1}^{m}\mathbf{u}_t\mathbf{u}_t = \mathrm{U}^{\mathrm{T}}\mathrm{U} = \mathrm{I}_{d\times d}$.

To prove $(ii)$, let $\mathbf{z}$ be an arbitrary unit vector and consider

$$\mathbf{z}^{\mathrm{T}}\mathrm{X}\mathbf{z} = \mathbf{z}^{\mathrm{T}}\mathrm{S}^2\mathbf{z} - \frac{1}{p}(\mathbf{z}^{\mathrm{T}}\mathrm{S}\mathbf{u})^2.$$

It follows that $\mathbf{z}^{\mathrm{T}}\mathrm{X}\mathbf{z} \leq \|\mathrm{S}\|^2$. To get a lower bound, we use $p \geq \beta\mathbf{u}^{\mathrm{T}}\mathrm{S}^2\mathbf{u}/\mathrm{trace}(\mathrm{S}^2)$:

$$\begin{aligned}\mathbf{z}^{\mathrm{T}}\mathrm{X}\mathbf{z} &\geq \mathbf{z}^{\mathrm{T}}\mathrm{S}^2\mathbf{z} - \frac{\mathrm{trace}(\mathrm{S}^2)}{\beta}\frac{(\mathbf{z}^{\mathrm{T}}\mathrm{S}\mathbf{u})^2}{\mathbf{u}^{\mathrm{T}}\mathrm{S}^2\mathbf{u}},\\ &\overset{(a)}{\geq} \|\mathrm{S}\|^2\left(\frac{\sigma_{\min}^2(\mathrm{S})}{\|\mathrm{S}\|^2} - \frac{\mathrm{trace}(\mathrm{S}^2)}{\beta\|S\|^2}\right),\\ &= \|\mathrm{S}\|^2\left(\frac{1}{\kappa^2} - \frac{\rho}{\beta}\right).\end{aligned}$$

(a) follows because: by definition of $\sigma_{\min}$, the minimum of the first term is $\sigma^2_{\min}$; and, by Cauchy-Schwarz, $(\mathbf{z}^{\mathrm{T}}\mathrm{S}\mathbf{u})^2 \le (\mathbf{z}^{\mathrm{T}}\mathbf{z})(\mathbf{u}^{\mathrm{T}}\mathrm{S}^2\mathbf{u})$. Since $\beta \le 1$, $\rho/\beta - \kappa^{-2} \ge 1$, and so $|\mathbf{z}^{\mathrm{T}}\mathrm{X}\mathbf{z}| \le \|\,\mathrm{S}\,\|^2\left(\rho/\beta - \kappa^{-2}\right)$, from which $(ii)$ follows.

To prove $(iii)$, first note that

$$\begin{aligned}
\mathbb{E}[\mathrm{X}^{\mathrm{T}}\mathrm{X}] &= \mathrm{S}^4 - \mathrm{S}^3\,\mathbb{E}\,[\mathbf{u}\mathbf{u}^{\mathrm{T}}/p]\mathrm{S} - \mathrm{S}\,\mathbb{E}\,[\mathbf{u}\mathbf{u}^{\mathrm{T}}/p]\mathrm{S}^3 + \mathrm{S}\,\mathbb{E}\,[\mathbf{u}\mathbf{u}^{\mathrm{T}}\mathrm{S}^2\mathbf{u}\mathbf{u}^{\mathrm{T}}/p^2]\mathrm{S}, \\
&\overset{(a)}{=} \mathrm{S}\left(\sum_{t=1}^{m}\frac{1}{p_t}\mathbf{u}_t\mathbf{u}_t^{\mathrm{T}}\mathrm{S}^2\mathbf{u}_t\mathbf{u}_t^{\mathrm{T}}\right)\mathrm{S} - \mathrm{S}^4.
\end{aligned}$$

(a) follows because $\mathbb{E}[\mathbf{u}\mathbf{u}^{\mathrm{T}}/p] = \mathrm{I}$. Thus, for an arbitrary unit $\mathbf{z}$, we have

$$\begin{aligned}
\mathbf{z}^{\mathrm{T}}\,\mathbb{E}\,[\mathrm{X}^{\mathrm{T}}\mathrm{X}]\mathbf{z} &= \sum_{t=1}^{m}\frac{1}{p_t}(\mathbf{z}^{\mathrm{T}}\mathrm{S}\mathbf{u}_t\mathbf{u}_t^{\mathrm{T}}\mathrm{S}\mathbf{z})\mathbf{u}_t^{\mathrm{T}}\mathrm{S}^2\mathbf{u}_t - \mathbf{z}^{\mathrm{T}}\mathrm{S}^4\mathbf{z}^{\mathrm{T}}, \\
&\overset{(a)}{\le} \frac{\mathrm{trace}(\mathrm{S}^2)}{\beta}\mathbf{z}^{\mathrm{T}}\mathrm{S}\left(\sum_{t=1}^{m}\mathbf{u}_t\mathbf{u}_t^{\mathrm{T}}\right)\mathrm{S}\mathbf{z} - \mathbf{z}^{\mathrm{T}}\mathrm{S}^4\mathbf{z}^{\mathrm{T}}, \\
&\overset{(b)}{=} \|\,\mathrm{S}\,\|^4\left(\frac{\mathrm{trace}(\mathrm{S}^2)}{\beta\|\,\mathrm{S}\,\|^2}\frac{\mathbf{z}^{\mathrm{T}}\mathrm{S}^2\mathbf{z}^{\mathrm{T}}}{\|\,\mathrm{S}\,\|^2} - \frac{\mathbf{z}^{\mathrm{T}}\mathrm{S}^4\mathbf{z}^{\mathrm{T}}}{\|\,\mathrm{S}\,\|^4}\right), \\
&\le \|\,\mathrm{S}\,\|^4\left(\frac{\mathrm{trace}(\mathrm{S}^2)}{\beta\|\,\mathrm{S}\,\|^2} - \frac{\sigma^4_{\min}}{\|\,\mathrm{S}\,\|^4}\right).
\end{aligned}$$

(a) follows from $p_t \ge \beta\mathbf{u}_t^{\mathrm{T}}\mathrm{S}^2\mathbf{u}_t/\mathrm{trace}(\mathrm{S}^2)$; (b) follows from $\mathrm{U}^{\mathrm{T}}\mathrm{U} = \sum_{t=1}^{m}\mathbf{u}_t\mathbf{u}_t^{\mathrm{T}} = \mathrm{I}_{d\times d}$. Thus, $|\mathbf{z}^{\mathrm{T}}\,\mathbb{E}\,[\mathrm{X}^{\mathrm{T}}\mathrm{X}]\mathbf{z}| \le \|\,\mathrm{S}\,\|^4(\rho/\beta - \kappa^{-4})$, from which $(iii)$ follows. ■

For the general case, consider two orthonormal matrices $\mathrm{W} \in \mathbb{R}^{m\times d_1}$, $\mathrm{V} \in \mathbb{R}^{m\times d_2}$, and two positive diagonal matrices $\mathrm{S}_1 \in \mathbb{R}^{d_1\times d_1}$ and $\mathrm{S}_2 \in \mathbb{R}^{d_2\times d_2}$. We consider the product $\mathrm{S}_1\mathrm{W}^{\mathrm{T}}\mathrm{V}\mathrm{S}_2$, which is approximated by the sampled product $\mathrm{S}_1\mathrm{W}^{\mathrm{T}}\mathrm{Q}^{\mathrm{T}}\mathrm{Q}\mathrm{V}\mathrm{S}_2$. Consider the sampling probabilities

$$p_t \ge \beta\frac{(\mathbf{u}_t^{\mathrm{T}}\mathrm{S}_1^2\mathbf{u}_t)^{1/2}(\mathbf{v}_t^{\mathrm{T}}\mathrm{S}_2^2\mathbf{v}_t)^{1/2}}{\sum_{t=1}^{m}(\mathbf{u}_t^{\mathrm{T}}\mathrm{S}_1^2\mathbf{u}_t)^{1/2}(\mathbf{v}_t^{\mathrm{T}}\mathrm{S}_2^2\mathbf{v}_t)^{1/2}} \ge \beta\frac{(\mathbf{u}_t^{\mathrm{T}}\mathrm{S}_1^2\mathbf{u}_t)^{1/2}(\mathbf{v}_t^{\mathrm{T}}\mathrm{S}_2^2\mathbf{v}_t)^{1/2}}{\sqrt{\mathrm{trace}(\mathrm{S}_1^2)\mathrm{trace}(\mathrm{S}_2^2)}},$$

where the last inequality follows from Cauchy-Schwarz. Since $\|\,\mathrm{A}\,\|_F = \sqrt{\rho(\mathrm{A})}\|\,A\,\| \ge \|\,A\,\|$, one immediately has from Drineas *et al.* (2006a) (using a simplified form for the bound),

$$\mathbb{P}\left[\|\,\mathrm{S}_1\mathrm{W}^{\mathrm{T}}\mathrm{V}\mathrm{S}_2 - \mathrm{S}_1\mathrm{W}^{\mathrm{T}}\mathrm{Q}^{\mathrm{T}}\mathrm{Q}\mathrm{V}\mathrm{S}_2\,\| > \epsilon\|\,\mathrm{S}_1\,\|\|\,\mathrm{S}_2\,\|\right] \le \exp\left(\frac{-r\beta^2\epsilon^2}{16\rho_1\rho_2}\right), \tag{3}$$

where $\rho_1 = \rho(\mathrm{S}_1)$ and $\rho_2 = \rho(\mathrm{S}_2)$. Alternatively, if $r \ge (16\rho_1\rho_2/\beta^2\epsilon^2)\ln\frac{1}{\delta}$, then

$$\|\,\mathrm{S}_1\mathrm{W}^{\mathrm{T}}\mathrm{V}\mathrm{S}_2 - \mathrm{S}_1\mathrm{W}^{\mathrm{T}}\mathrm{Q}^{\mathrm{T}}\mathrm{Q}\mathrm{V}\mathrm{S}_2\,\| \le \epsilon\|\,\mathrm{S}_1\,\|\|\,\mathrm{S}_2\,\|.$$

The dependence on the stable ranks and $\beta$ is quadratic. Applying this bound to the situation in Lemma 15 would give an inferior bound. The intuition behind the improvement is that the sampling is isotropic, and so will not favor any particular direction. One can therefore guess that all the singular values are approximately equal and so the Frobenius norm bound on the spectral norm will be loose by a factor of $\sqrt{\rho}$; and, indeed this is what comes out in the closer analysis. As a application of Lemma 15, we can get a result for the asymmetric case.

**Lemma 17.** *Let* $\mathrm{W} \in \mathbb{R}^{m\times d_1}$, $\mathrm{V} \in \mathbb{R}^{m\times d_2}$ *be orthonormal, and let* $\mathrm{S}_1 \in \mathbb{R}^{d_1\times d_1}$ *and* $\mathrm{S}_2 \in \mathbb{R}^{d_2\times d_2}$ *be two positive diagonal matrices. Consider row sampling probabilities*

$$p_t \geq \beta \frac{\frac{1}{\|\mathrm{S}_1\|^2}\mathbf{w}_t^{\mathrm{T}}\mathrm{S}_1^2\mathbf{w}_t + \frac{1}{\|\mathrm{S}_2\|^2}\mathbf{v}_t^{\mathrm{T}}\mathrm{S}_2^2\mathbf{v}_t}{\rho_1+\rho_2}.$$

*If* $r \geq (8(\rho_1+\rho_2)/\beta\epsilon^2)\ln\frac{2(d_1+d_2)}{\delta}$, *then with probability at least* $1-\delta$,

$$\|\mathrm{S}_1\mathrm{W}^{\mathrm{T}}\mathrm{V}\mathrm{S}_2 - \mathrm{S}_1\mathrm{W}^{\mathrm{T}}\mathrm{Q}^{\mathrm{T}}\mathrm{Q}\mathrm{V}\mathrm{S}_2\| \leq \epsilon\|\mathrm{S}_1\|\|\mathrm{S}_2\|$$

For the special case that $\mathrm{S}_1 = \mathrm{I}_{d_1\times d_1}$ and $\mathrm{S}_2 = \mathrm{I}_{d_2\times d_2}$, the sampling probabilities simplify to

$$p_t \geq \beta\frac{\mathbf{w}_t^{\mathrm{T}}\mathbf{w}_t + \mathbf{v}_t^{\mathrm{T}}\mathbf{v}_t}{d_1+d_2},$$

**Corollary 18.** *If* $r \geq (8(d_1+d_2)/\beta\epsilon^2)\ln\frac{2(d_1+d_2)}{\delta}$, *then with probability at least* $1-\delta$,

$$\|\mathrm{W}^{\mathrm{T}}\mathrm{V} - \mathrm{W}^{\mathrm{T}}\mathrm{Q}^{\mathrm{T}}\mathrm{Q}\mathrm{V}\| \leq \epsilon.$$

*Proof.* (of Lemma 17) By homogeneity, we can without loss of generality assume that $\|\mathrm{S}_1\| = \|\mathrm{S}_2\| = 1$, and let[1] $\mathrm{Z} = [\mathrm{W}\mathrm{S}_1\ \mathrm{V}\mathrm{S}_2]$. An elementary lemma which we will find useful is

**Lemma 19.** *For any matrix* $\mathrm{A} = [\mathrm{A}_1\ \mathrm{A}_2]$,

$$\max\{\|\mathrm{A}_1\|, \|\mathrm{A}_2\|\} \leq \|\mathrm{A}\| \leq \sqrt{\|\mathrm{A}_1\|^2 + \|\mathrm{A}_2\|^2}.$$

The left inequality is saturated when $\mathrm{A}_1$ and $\mathrm{A}_2$ are orthogonal ($\mathrm{A}_1^{\mathrm{T}}\mathrm{A}_2 = \mathbf{0}$), and the right hand inequality is saturated when $\mathrm{A}_1 = \mathrm{A}_2$. By repeatedly applying Lemma 19 one can see that $\|\mathrm{A}\|$ is at least the spectral norm of any submatrix. Introduce the SVD of Z,

$$\mathrm{Z} = [\mathrm{W}\mathrm{S}_1\ \mathrm{V}\mathrm{S}_2] = \mathrm{U}\mathrm{S}\mathrm{V}^{\mathrm{T}}.$$

We use our typical row sampling probabilities according to US,

$$p_t \geq \beta\frac{\mathbf{u}_t^{\mathrm{T}}\mathrm{S}^2\mathbf{u}_t}{\mathrm{trace}(\mathrm{S}^2)}.$$

We may interpret the sampling probabilities as follows. Let $\mathbf{z}_t$ be a row of Z, the concatenation of two rows in $\mathrm{W}\mathrm{S}_1$ and $\mathrm{V}\mathrm{S}_2$: $\mathbf{z}_t^{\mathrm{T}} = [\mathbf{w}_t^{\mathrm{T}}\mathrm{S}_1\ \mathbf{v}_t^{\mathrm{T}}\mathrm{S}_2]$. We also have that $\mathbf{z}_t^{\mathrm{T}} = \mathbf{u}_t^{\mathrm{T}}\mathrm{S}$. Hence,

$$\mathbf{u}_t^{\mathrm{T}}\mathrm{S}^2\mathbf{u}_t = \mathbf{z}_t^{\mathrm{T}}\mathbf{z}_t = \mathbf{w}_t^{\mathrm{T}}\mathrm{S}_1^2\mathbf{w}_t + \mathbf{v}_t^{\mathrm{T}}\mathrm{S}_2^2\mathbf{v}_t.$$

These are exactly the probabilities as claimed in the statement of the lemma (modulo the rescaling).

Applying Lemma 15: if $r \geq (4\rho/\beta\epsilon^2)\ln\frac{2\cdot\mathrm{rank}(\mathrm{U})}{\delta}$, then with probability at least $1-\delta$,

$$\|\mathrm{S}^2 - \mathrm{S}\mathrm{U}^{\mathrm{T}}\mathrm{Q}^{\mathrm{T}}\mathrm{Q}\mathrm{U}\mathrm{S}\| \leq \epsilon\|\mathrm{S}\|^2 \leq \epsilon\sqrt{\|\mathrm{S}_1\|^2 + \|\mathrm{S}_2\|^2} = \epsilon\sqrt{2},$$

where the second inequality follows from Lemma 19. Since ZV = US,

$$\|\mathrm{Z}^{\mathrm{T}}\mathrm{Z} - \mathrm{Z}^{\mathrm{T}}\mathrm{Q}^{\mathrm{T}}\mathrm{Q}\mathrm{Z}\| = \|\mathrm{S}^2 - \mathrm{S}\mathrm{U}^{\mathrm{T}}\mathrm{Q}^{\mathrm{T}}\mathrm{Q}\mathrm{U}\mathrm{S}\|.$$

[1] The general case would have been $\mathrm{Z} = \left[\frac{1}{\|\mathrm{S}_1\|}\mathrm{W}\mathrm{S}_1\ \frac{1}{\|\mathrm{S}_2\|}\mathrm{V}\mathrm{S}_2\right]$.

Further, by the construction of Z,

$$\mathrm{Z^TZ - Z^TQ^TQZ} = \begin{bmatrix} \mathrm{S_1^2 - S_1W^TQ^TQWS_1} & \mathrm{S_1W^TVS_2 - S_1W^TQ^TQVS_2} \\ \mathrm{S_2V^TWS_1 - S_2V^TQ^TQWS_1} & \mathrm{S_2^2 - S_2V^TQ^TQVS_2} \end{bmatrix}.$$

By Lemma 19, $\| \mathrm{S_1W^TVS_2 - S_1W^TQ^TQVS_2} \| \leq \| \mathrm{Z^TZ - Z^TQ^TQZ} \|$, and so:

$$\| \mathrm{S_1W^TVS_2 - S_1W^TQ^TQVS_2} \| \leq \epsilon\sqrt{2}.$$

Observe that $\mathrm{trace}(\mathrm{S}^2) = \| \mathrm{Z} \|_F^2 = \mathrm{trace}(\mathrm{S}_1^2) + \mathrm{trace}(\mathrm{S}_2^2)$; further, since $\| \mathrm{S} \| \geq \max\{\| \mathrm{S}_1 \|, \| \mathrm{S}_2 \|\}$, we have that

$$\rho(\mathrm{S}) = \frac{\mathrm{trace}(\mathrm{S}^2)}{\| \mathrm{S} \|^2} = \frac{\mathrm{trace}(\mathrm{S}_1^2) + \mathrm{trace}(\mathrm{S}_2^2)}{\| \mathrm{S} \|^2} \leq \frac{\mathrm{trace}(\mathrm{S}_1^2)}{\| \mathrm{S}_1 \|^2} + \frac{\mathrm{trace}(\mathrm{S}_2^2)}{\| \mathrm{S}_2 \|^2} = \rho_1 + \rho_2.$$

Since $\mathrm{rank}(\mathrm{U}) \leq d_1 + d_2$, it suffices that $r \geq (4(\rho_1 + \rho_2)/\beta\epsilon^2) \ln \frac{2(d_1+d_2)}{\delta}$ to obtain error $\epsilon\sqrt{2}$; after rescaling $\epsilon' = \epsilon\sqrt{2}$, we have the result. ■

# 4 Sampling for Matrix Multiplication

We obtain results for matrix multiplication directly from Lemmas 15 and 17. First we consider the symmetric case, then the asymmetric case. Let $\mathrm{A} \in \mathbb{R}^{m \times d_1}$ and $\mathrm{B} \in \mathbb{R}^{m \times d_2}$. We are interested in conditions on the sampling matrix $\mathrm{Q} \in \mathbb{R}^{r \times m}$ such that $\mathrm{A^TA} \approx \tilde{\mathrm{A}}^{\mathrm{T}}\tilde{\mathrm{A}}$ and $\mathrm{A^TB} \approx \tilde{\mathrm{A}}^{\mathrm{T}}\tilde{\mathrm{B}}$, where $\tilde{\mathrm{A}} = \mathrm{QA}$ and $\tilde{\mathrm{B}} = \mathrm{QB}$. Using the SVD of A,

$$\begin{aligned} \| \mathrm{A^TA - A^TQ^TQA} \| &= \| \mathrm{V_AS_AU_A^TU_AS_AV_A^T - V_AS_AU_A^TQ^TQU_AS_AV_A^T} \|, \\ &= \| \mathrm{S_A^2 - S_AU_A^TQ^TQU_AS_A} \|. \end{aligned}$$

We may now directly apply Lemma 15, with respect to the appropriate sampling probabilities. One can verify that the sampling probabilities in Lemma 15 are proportional to the squared norms of the rows of A.

**Theorem 20.** *Let* $\mathrm{A} \in \mathbb{R}^{m \times d_1}$ *have rows* $\mathbf{a}_t$ *Obtain a sampling matrix* $\mathrm{Q} \in \mathbb{R}^{r \times m}$ *using row-sampling probabilities*

$$p_t \geq \beta \frac{\mathbf{a}_t^{\mathrm{T}}\hat{\mathbf{a}}_t}{\| \mathrm{A} \|_F^2}.$$

*Then, if* $r \geq \frac{4\rho_{\mathrm{A}}}{\beta\epsilon^2} \ln \frac{2d_1}{\delta}$, *with probability at least* $1 - \delta$,

$$\| \mathrm{A^TA} - \tilde{\mathrm{A}}^{\mathrm{T}}\tilde{\mathrm{A}} \| \leq \epsilon \| \mathrm{A} \|^2.$$

Similarly, using the SVDs of A and B,

$$\begin{aligned} \| \mathrm{A^TB - A^TQ^TQB} \| &= \| \mathrm{V_AS_AU_A^TU_BS_BV_B^T - V_AS_AU_A^TQ^TQU_BS_BV_B^T} \|, \\ &= \| \mathrm{S_AU_A^TU_BS_B - S_AU_A^TQ^TQU_BS_B} \|. \end{aligned}$$

We may now directly apply Lemma 17, with respect to the appropriate sampling probabilities. One can verify that the sampling probabilities in Lemma 17 are proportional to the sum of the rescaled squared norms of the rows of A and B.

**Theorem 21.** *Let* $\mathrm{A} \in \mathbb{R}^{m \times d_1}$ *and* $\mathrm{B} \in \mathbb{R}^{m \times d_2}$, *have rescaled rows* $\hat{\mathbf{a}}_t = \mathbf{a}_t / \|\mathrm{A}\|$ *and* $\hat{\mathbf{b}}_t = \mathbf{b}_t \|\mathrm{B}\|$ *respectively. Obtain a sampling matrix* $\mathrm{Q} \in \mathbb{R}^{r \times m}$ *using row-sampling probabilities*

$$p_t \geq \beta \frac{\hat{\mathbf{a}}_t^{\mathrm{T}} \hat{\mathbf{a}}_t + \hat{\mathbf{b}}_t^{\mathrm{T}} \hat{\mathbf{b}}_t}{\sum_{t=1}^m \hat{\mathbf{a}}_t^{\mathrm{T}} \hat{\mathbf{a}}_t + \hat{\mathbf{b}}_t^{\mathrm{T}} \hat{\mathbf{b}}_t} = \beta \frac{\hat{\mathbf{a}}_t^{\mathrm{T}} \hat{\mathbf{a}}_t + \hat{\mathbf{b}}_t^{\mathrm{T}} \hat{\mathbf{b}}_t}{\rho_{\mathrm{A}} + \rho_{\mathrm{B}}}.$$

*Then, if* $r \geq \frac{8(\rho_{\mathrm{A}} + \rho_{\mathrm{B}})}{\beta \epsilon^2} \ln \frac{2(d_1 + d_2)}{\delta}$, *with probability at least* $1 - \delta$,

$$\|\mathrm{A}^{\mathrm{T}}\mathrm{B} - \tilde{\mathrm{A}}^{\mathrm{T}}\tilde{\mathrm{B}}\| \leq \epsilon \|\mathrm{A}\| \|\mathrm{B}\|.$$

# 5 Sparse Row Based Matrix Representation

Given a matrix $\mathrm{A} = \mathrm{USV}^{\mathrm{T}} \in \mathbb{R}^{m \times d}$, the top $k$ singular vectors, corresponding to the top $k$ singular values give the best rank $k$ reconstruction of A. Specifically, let $\mathrm{A}_k = \mathrm{U}_k \mathrm{S}_k \mathrm{V}_k^{\mathrm{T}}$, where $\mathrm{U}_k \in \mathbb{R}^{m \times k}$, $\mathrm{S}_k \in \mathbb{R}^{k \times k}$ and $\mathrm{V}_k \in \mathbb{R}^{d \times k}$; then, $\|\mathrm{A} - \mathrm{A}_k\| \leq \|\mathrm{A} - \mathrm{X}\|$ where $\mathrm{X} \in \mathbb{R}^{m \times d}$ ranges over all rank-$k$ matrices. As usual, let $\tilde{\mathrm{A}} = \mathrm{QA}$ be the sampled, rescaled rows of A, with $\tilde{\mathrm{A}} = \tilde{\mathrm{U}}\tilde{\mathrm{S}}\tilde{\mathrm{V}}^{\mathrm{T}}$, and consider the top-$k$ right singular vectors $\tilde{\mathrm{V}}_k$. Let $\tilde{\Pi}_k$ be the projection onto this top-$k$ right singular space, and consider the rank $k$ approximation to A obtained by projecting onto this space: $\tilde{\mathrm{A}}_k = \mathrm{A}\tilde{\Pi}_k$. The following lemma is useful for showing that $\tilde{\mathrm{A}}_k$ is almost (up to additive error) as good an approximation to A as one can get.

**Lemma 22** (Drineas *et al.* (2006b), Rudelson and Vershynin (2007))**.**

$$\|\mathrm{A} - \tilde{\mathrm{A}}_k\|^2 \leq \|\mathrm{A} - \mathrm{A}_k\|^2 + 2\|\mathrm{A}^{\mathrm{T}}\mathrm{A} - \tilde{\mathrm{A}}^{\mathrm{T}}\tilde{\mathrm{A}}\| \leq (\|\mathrm{A} - \mathrm{A}_k\| + \sqrt{2}\|\mathrm{A}^{\mathrm{T}}\mathrm{A} - \tilde{\mathrm{A}}^{\mathrm{T}}\tilde{\mathrm{A}}\|^{1/2})^2.$$

*Proof.* The proof follows using standard arguments and an application of a perturbation theory result due to Weyl for bounding the change in any singular value upon hermitian perturbation of a hermitian matrix. ■

Therefore, if we can approximate the matrix product $\mathrm{A}^{\mathrm{T}}\mathrm{A}$, we immediately get a good reconstruction for every $k$. The appropriate sampling probabilities from the previous section are

$$p_t \geq \beta \frac{\mathbf{a}_t^{\mathrm{T}} \mathbf{a}_t}{\|\mathrm{A}\|_F^2}.$$

In this case, if $r \geq (4\rho/\beta\epsilon^2) \ln \frac{2d}{\delta}$, then with probability at least $1 - \delta$,

$$\|\mathrm{A} - \tilde{\mathrm{A}}_k\|^2 \leq \|\mathrm{A} - \mathrm{A}_k\|^2 + 2\epsilon \|\mathrm{A}\|^2.$$

The sampling probabilities are easy to compute and sampling can be accomplished in one pass if the matrix is stored row-by-row.

To get a relative error result, we need a more careful non-uniform sampling probabilities. The problem here becomes apparent if A has rank $k$. In this case we have no hope of a relative error approximation unless we preserve the rank during sampling. To do so, we need to sample according to the actual singular vectors in U, not according to A; this is because sampling according to A can give especially large weight to a few of the large singular value directions, ignoring the small singular value directions and hence not preserving rank. By sampling according to U, we essentially put equal weight on all singular directions. To approximate U well, we need sampling probabilities

$$p_t \geq \frac{\beta}{d} \mathbf{u}_t^{\mathrm{T}} \mathbf{u}_t.$$

Then, from Corollary 16, if $r \geq (4(d-\beta)/\beta\epsilon^2)\ln\frac{2d}{\delta}$, with probability at least $1-\delta$,

$$\| \mathrm{I} - \mathrm{U}^{\mathrm{T}}\mathrm{Q}^{\mathrm{T}}\mathrm{Q}\mathrm{U} \| \leq \epsilon.$$

Since $\| \mathrm{U} \| = 1$, it also follows that

$$\| \mathrm{U}\mathrm{U}^{\mathrm{T}} - \mathrm{U}\mathrm{U}^{\mathrm{T}}\mathrm{Q}^{\mathrm{T}}\mathrm{Q}\mathrm{U}\mathrm{U}^{\mathrm{T}} \| \leq \epsilon.$$

This result is useful because of the following lemma.

**Lemma 23** (Spielman and Srivastava (2008))**.** *If* $\| \mathrm{U}\mathrm{U}^{\mathrm{T}} - \mathrm{U}\mathrm{U}^{\mathrm{T}}\mathrm{Q}^{\mathrm{T}}\mathrm{Q}\mathrm{U}\mathrm{U}^{\mathrm{T}} \| \leq \epsilon$*, then for every* $\mathbf{x} \in \mathbb{R}^d$*,*

$$(1-\epsilon)\mathbf{x}^{\mathrm{T}}\mathrm{A}^{\mathrm{T}}\mathrm{A}\mathbf{x} \leq \mathbf{x}^{\mathrm{T}}\tilde{\mathrm{A}}^{\mathrm{T}}\tilde{\mathrm{A}}\mathbf{x} \leq (1+\epsilon)\mathbf{x}^{\mathrm{T}}\mathrm{A}^{\mathrm{T}}\mathrm{A}\mathbf{x}.$$

*Proof.* We give a sketch of the proof from Spielman and Srivastava (2008). We let $\mathbf{x} \neq \mathbf{0}$ range over col(U). Since col(U) = col(A), $\mathbf{x} \in$ col(U) if and only if for some $\mathbf{y} \in \mathbb{R}^d$, $\mathbf{x} = \mathrm{A}\mathbf{y}$. Since rank(A) $= d$, $\mathrm{A}\mathbf{y} \neq 0 \iff \mathbf{y} \neq 0$. Also note that $\mathrm{U}\mathrm{U}^{\mathrm{T}}\mathrm{A} = \mathrm{A}$, since $\mathrm{U}\mathrm{U}^{\mathrm{T}}$ is a projection operator onto the column space of U, which is the same as the column space of A. The following sequence establishes the lemma.

$$\begin{aligned}
\| \mathrm{U}\mathrm{U}^{\mathrm{T}} - \mathrm{U}\mathrm{U}^{\mathrm{T}}\mathrm{Q}^{\mathrm{T}}\mathrm{Q}\mathrm{U}\mathrm{U}^{\mathrm{T}} \| &= \sup_{\mathbf{x}\neq\mathbf{0}} \frac{|\mathbf{x}^{\mathrm{T}}\mathrm{U}\mathrm{U}^{\mathrm{T}}\mathbf{x} - \mathbf{x}^{\mathrm{T}}\mathrm{U}\mathrm{U}^{\mathrm{T}}\mathrm{Q}^{\mathrm{T}}\mathrm{Q}\mathrm{U}\mathrm{U}^{\mathrm{T}}\mathbf{x}|}{\mathbf{x}^{\mathrm{T}}\mathbf{x}}, \\
&= \sup_{\mathrm{A}\mathbf{y}\neq\mathbf{0}} \frac{|\mathbf{y}^{\mathrm{T}}\mathrm{A}^{\mathrm{T}}\mathrm{U}\mathrm{U}^{\mathrm{T}}\mathrm{A}\mathbf{y} - \mathbf{y}^{\mathrm{T}}\mathrm{A}^{\mathrm{T}}\mathrm{U}\mathrm{U}^{\mathrm{T}}\mathrm{Q}^{\mathrm{T}}\mathrm{Q}\mathrm{U}\mathrm{U}^{\mathrm{T}}\mathrm{A}\mathbf{y}|}{\mathbf{y}^{\mathrm{T}}\mathrm{A}^{\mathrm{T}}\mathrm{A}\mathbf{y}}, \\
&= \sup_{\mathrm{A}\mathbf{y}\neq\mathbf{0}} \frac{|\mathbf{y}^{\mathrm{T}}\mathrm{A}^{\mathrm{T}}\mathrm{A}\mathbf{y} - \mathbf{y}^{\mathrm{T}}\mathrm{A}^{\mathrm{T}}\mathrm{Q}^{\mathrm{T}}\mathrm{Q}\mathrm{A}\mathbf{y}|}{\mathbf{y}^{\mathrm{T}}\mathrm{A}^{\mathrm{T}}\mathrm{A}\mathbf{y}}, \\
&= \sup_{\mathbf{y}\neq\mathbf{0}} \frac{|\mathbf{y}^{\mathrm{T}}\mathrm{A}^{\mathrm{T}}\mathrm{A}\mathbf{y} - \mathbf{y}^{\mathrm{T}}\tilde{\mathrm{A}}^{\mathrm{T}}\tilde{\mathrm{A}}\mathbf{y}|}{\mathbf{y}^{\mathrm{T}}\mathrm{A}^{\mathrm{T}}\mathrm{A}\mathbf{y}},
\end{aligned}$$

The lemma now follows because $\| \mathrm{U}\mathrm{U}^{\mathrm{T}} - \mathrm{U}\mathrm{U}^{\mathrm{T}}\mathrm{Q}^{\mathrm{T}}\mathrm{Q}\mathrm{U}\mathrm{U}^{\mathrm{T}} \| \leq \epsilon$. ∎

Via the Courant-Fischer characterization of the singular values, it is immediate from Lemma 23 that the singular value spectrum is also preserved :

$$(1-\epsilon)\sigma_i(\mathrm{A}^{\mathrm{T}}\mathrm{A}) \leq \sigma_i(\tilde{\mathrm{A}}^{\mathrm{T}}\tilde{\mathrm{A}}) \leq (1+\epsilon)\sigma_i(\mathrm{A}^{\mathrm{T}}\mathrm{A}). \tag{4}$$

Lemma 23 along with (4) will allow us to prove the relative approximation result.

**Theorem 24.** *If* $p_t \geq \frac{\beta}{d}\mathbf{u}_t^{\mathrm{T}}\mathbf{u}_t$ *and* $r \geq (4(d-\beta)/\beta\epsilon^2)\ln\frac{2d}{\epsilon}$*, then, for* $k = 1, \ldots, d$*,*

$$\| \mathrm{A} - \mathrm{A}\tilde{\Pi}_k \| \leq \left(\frac{1+\epsilon}{1-\epsilon}\right)^{1/2} \| \mathrm{A} - \mathrm{A}_k \|,$$

*where* $\tilde{\Pi}_k$ *projects onto the top* $k$ *right singular vectors of* $\tilde{\mathrm{A}}$*.*

**Remarks.** For $\epsilon \le \frac{1}{2}$, $\left(\frac{1+\epsilon}{1-\epsilon}\right)^{1/2} \le 1 + 2\epsilon$. Computing the probabilities $p_t$ involves knowing $\mathbf{u}_t$ which means one has to perform an $SVD$, in which case, one could use $\mathrm{A}_k$; it seems like overkill to compute $\mathrm{A}_k$ in order to approximate $\mathrm{A}_k$. We discuss approximate sampling schemes later, in Section 7.

*Proof.* Let $\| \mathbf{x} \| = 1$. The following sequence establishes the result.

$$\begin{aligned}
\| \mathrm{A}(\mathrm{I} - \tilde{\Pi}_k) \|^2 &= \sup_{\mathbf{x}\in\ker(\tilde{\Pi}_k)} \| \mathrm{A}\mathbf{x} \|^2 = \sup_{\mathbf{x}\in\ker(\tilde{\Pi}_k)} \mathbf{x}^{\mathrm{T}}\mathrm{A}^{\mathrm{T}}\mathrm{A}\mathbf{x},\\
&\le \frac{1}{1-\epsilon} \sup_{\mathbf{x}\in\ker(\tilde{\Pi}_k)} \mathbf{x}^{\mathrm{T}}\tilde{\mathrm{A}}^{\mathrm{T}}\tilde{\mathrm{A}}\mathbf{x},\\
&= \frac{1}{1-\epsilon}\sigma_{k+1}(\tilde{\mathrm{A}}^{\mathrm{T}}\tilde{\mathrm{A}}),\\
&\le \frac{1+\epsilon}{1-\epsilon}\sigma_{k+1}(\mathrm{A}^{\mathrm{T}}\mathrm{A}) = \frac{1+\epsilon}{1-\epsilon}\| \mathrm{A} - \mathrm{A}_k \|^2.
\end{aligned}$$

■

# 6 $\ell_2$ Linear Regression with Relative Error Bounds

A linear regression is represented by a real data matrix $\mathrm{A} \in \mathbb{R}^{m\times d}$ which represents $m$ points in $\mathbb{R}^d$, and a target vector $\mathbf{y} \in \mathbb{R}^m$. Traditionally, $m \gg d$ (severly over constrained regression). The goal is to find a regression vector $\mathbf{x}^* \in \mathbb{R}^2$ which minimizes the $\ell_2$ fit error (least squares regression)

$$\mathcal{E}(\mathbf{x}) = \| \mathrm{A}\mathbf{x} - \mathbf{y} \|_2^2 = \sum_{t=1}^{m}(\mathbf{a}_t^{\mathrm{T}}\mathbf{x} - y_t)^2,$$

We assume such an optimal $\mathbf{x}^*$ exists (it may not be unique unless A has full column rank), and is given by $\mathbf{x}^* = \mathrm{A}^+\mathbf{y}$, where $^+$ denotes the More-Penrose pseudo-inverse; this problem can be solved in $O(md^2)$. Through row-sampling, it is possible to construct $\hat{\mathbf{x}}$, an approximation to the optimal regression weights $\mathbf{x}^*$, which is a relative error approximation to optimal,

$$\mathcal{E}(\hat{\mathbf{x}}) \le (1+\epsilon)\mathcal{E}(\mathbf{x}^*).$$

As usual, let $\mathrm{A} = \mathrm{U_A S_A V_A^T}$. Then $\mathrm{A}^+ = \mathrm{V_A S_A^{-1} U_A^T}$, and so $\mathbf{x}^* = \mathrm{VS^{-1}U^T}\mathbf{y}$. The predictions are $\mathbf{y}^* = \mathrm{A}\mathbf{x}^* = \mathrm{U_A U_A^T}\mathbf{y}$, which is the projection of $\mathbf{y}$ onto the column space of A. We define the residual $\boldsymbol{\epsilon} = \mathbf{y} - \mathbf{y}^* = \mathbf{y} - \mathrm{A}\mathbf{x}^* = (\mathrm{I} - \mathrm{U_A U_A^T})\mathbf{y}$, so

$$\mathbf{y} = \mathrm{U_A U_A^T}\mathbf{y} + \boldsymbol{\epsilon}. \tag{5}$$

We will construct $\tilde{\mathrm{A}}$ and $\tilde{\mathbf{y}}$ by sampling rows:

$$[\tilde{\mathrm{A}}, \tilde{\mathbf{y}}] = \mathrm{Q}[\mathrm{A}, \mathbf{y}],$$

and solve the linear regression problem on $(\tilde{\mathrm{A}}, \tilde{\mathbf{y}})$ to obtain $\hat{\mathbf{x}} = \tilde{\mathrm{A}}^+\tilde{\mathbf{y}}$. For $\beta \in (0, \frac{1}{3}]$, we will use the sampling probabilities

$$p_t \ge \beta\left(\frac{\mathbf{u}_t^2}{d} + \frac{(\mathbf{u}_t^2 + \frac{\epsilon_t^2}{\boldsymbol{\epsilon}^{\mathrm{T}}\boldsymbol{\epsilon}})}{d+1} + \frac{\epsilon_t^2}{\boldsymbol{\epsilon}^{\mathrm{T}}\boldsymbol{\epsilon}}\right) \tag{6}$$

to construct $\tilde{\mathrm{A}}$ and $\tilde{\mathbf{y}}$. There are three parts to these sampling probabilities. The first part allows us to reconstruct A well from $\tilde{\mathrm{A}}$; the second allows us to reconstruct $\mathrm{A}^{\mathrm{T}}\boldsymbol{\epsilon}$; and, the third allows us to reconstruct $\boldsymbol{\epsilon}$.

Note that $\tilde{\mathrm{A}} = \mathrm{QU_A S_A V_A}^{\mathrm{T}}$; if $\mathrm{QU_A}$ consisted of orthonormal columns, then this would be the SVD of $\tilde{\mathrm{A}}$. Indeed, this is approximately so, as we will soon see. Let the SVD of $\tilde{\mathrm{A}}$ be $\tilde{\mathrm{A}} = \mathrm{U}_{\tilde{\mathrm{A}}}\mathrm{S}_{\tilde{\mathrm{A}}}\mathrm{V}_{\tilde{\mathrm{A}}}{}^{\mathrm{T}}$. Let $\tilde{\mathrm{U}} = \mathrm{QU_A}$. Since $p_t \geq \beta \mathbf{u}_t^2/d$, it follows from Corollary 16 that if $r \geq 2\frac{d-\beta}{\beta\epsilon^2}$, for $\epsilon \in (0,1)$, then, with high probability,

$$\| \mathrm{I} - \tilde{\mathrm{U}}^{\mathrm{T}}\tilde{\mathrm{U}} \| \leq \epsilon.$$

Since the eigenvalues of $\mathrm{I} - \tilde{\mathrm{U}}^{\mathrm{T}}\tilde{\mathrm{U}}$ are given by $1 - \sigma_i^2(\tilde{\mathrm{U}})$, it follows that

$$1 - \epsilon < \sigma_i^2(\tilde{\mathrm{U}}) < 1 + \epsilon.$$

So all the singular values of $\mathrm{U_A}$ are preserved after sampling. Essentially, it suffices to sample $r = O(d \ln d/\epsilon^2)$ rows to preserve the entire spectrum of $\mathrm{U_A}$. By choosing (say) $\epsilon = \frac{1}{2}$, the rank of $\mathrm{U_A}$ is preserved with high probability, since all the singular values are bigger than $\frac{1}{2}$. Thus, with high probability, $\mathrm{rank}(\tilde{\mathrm{A}}) = \mathrm{rank}(\mathrm{U}_{\tilde{\mathrm{A}}}) = \mathrm{rank}(\mathrm{QU_A}) = \mathrm{rank}(\mathrm{U_A}) = \mathrm{rank}(\mathrm{A})$. Since $\mathrm{QU_A}$ has full rank, $\mathrm{S}_{\mathrm{QU_A}}^{-1}$ is defined, and $\mathrm{S}_{\mathrm{QU_A}} - \mathrm{S}_{\mathrm{QU_A}}^{-1}$ is a diagonal matrix whose diagonals are $(\sigma_i^2(\tilde{\mathrm{U}}) - 1)/\sigma_i(\tilde{\mathrm{U}})$; thus, $\| \mathrm{S}_{\mathrm{QU_A}} - \mathrm{S}_{\mathrm{QU_A}}^{-1} \|_2 \leq \epsilon/\sqrt{1-\epsilon}$. This allows us to quantify the degree to which $\mathrm{QU_A}$ is orthonormal, because

$$\begin{aligned} \| (\mathrm{QU_A})^+ - (\mathrm{QU_A})^{\mathrm{T}} \|_2 &= \| \mathrm{V}_{\mathrm{QU_A}}\mathrm{S}_{\mathrm{QU_A}}^{-1}\mathrm{U}_{\mathrm{QU_A}}{}^{\mathrm{T}} - \mathrm{V}_{\mathrm{QU_A}}\mathrm{S}_{\mathrm{QU_A}}\mathrm{U}_{\mathrm{QU_A}}^{\mathrm{T}} \|_2 \\ &= \| \mathrm{S}_{\mathrm{QU_A}}^{-1} - \mathrm{S}_{\mathrm{QU_A}} \|_2 \leq \frac{\epsilon}{\sqrt{1-\epsilon}}. \end{aligned}$$

Finally, we can get a convenient form for $\tilde{\mathrm{A}}^+ = (\mathrm{QA})^+$, because $\mathrm{QA} = \mathrm{QU_A S_A V_A^T}$ has full rank, and so $\mathrm{QU_A} = \mathrm{U}_{\mathrm{QU_A}}\mathrm{S}_{\mathrm{QU_A}}\mathrm{V}_{\mathrm{QU_A}}^{\mathrm{T}}$ has full rank (and hence is the product of full rank matrices). Thus,

$$\begin{aligned} (\mathrm{QA})^+ &= (\mathrm{U}_{\mathrm{QU_A}}\mathrm{S}_{\mathrm{QU_A}}\mathrm{V}_{\mathrm{QU_A}}^{\mathrm{T}}\mathrm{S_A V_A^T})^+, \\ &= \mathrm{V_A}(\mathrm{S}_{\mathrm{QU_A}}\mathrm{V}_{\mathrm{QU_A}}^{\mathrm{T}}\mathrm{S_A})^+\mathrm{U}_{\mathrm{QU_A}}^{\mathrm{T}}, \\ &= \mathrm{V_A S_A^{-1} V_{QU_A} S_{QU_A}^{-1} U_{QU_A}^T}, \\ &= \mathrm{V_A S_A^{-1}}(\mathrm{QU_A})^+, \end{aligned}$$

We summarize all this information in the next lemma.

**Lemma 25.** *If* $r \geq (4d/\beta\epsilon^2)\ln\frac{2d}{\delta}$, *with probability at least* $1-\delta$, *all of the following hold:*

$$\mathrm{rank}(\tilde{\mathrm{A}}) = \mathrm{rank}(\mathrm{U}_{\tilde{\mathrm{A}}}) = \mathrm{rank}(\mathrm{QU_A}) = \mathrm{rank}(\mathrm{U_A}) = \mathrm{rank}(\mathrm{A}); \tag{7}$$

$$\| \mathrm{S}_{\mathrm{QU_A}} - \mathrm{S}_{\mathrm{QU_A}}^{-1} \|_2 \leq \epsilon/\sqrt{1-\epsilon}; \tag{8}$$

$$\| (\mathrm{QU_A})^+ - (\mathrm{QU_A})^{\mathrm{T}} \|_2 \leq \epsilon/\sqrt{1-\epsilon}; \tag{9}$$

$$(\mathrm{QA})^+ = \mathrm{V_A S_A^{-1}}(\mathrm{QU_A})^+. \tag{10}$$

In Lemma 25 we have simplified the constant to 4; this is a strengthened form of Lemma 4.1 in Drineas *et al.* (2006d); in particular, the dependence on $d$ is near-linear.

Remember that $\hat{\mathbf{x}} = \tilde{\mathrm{A}}^{+}\tilde{\mathbf{y}}$; we now bound $\| A\hat{\mathbf{x}} - \mathbf{y} \|^2$. We only sketch the derivation which basically follows the line of reasoning in Drineas *et al.* (2006d). Under the conditions of Lemma 25, with probability at least $1 - \delta$,

$$\begin{aligned}
\| \mathrm{A}\hat{\mathbf{x}} - \mathbf{y} \| &= \| \mathrm{A}\tilde{\mathrm{A}}^{+}\tilde{\mathbf{y}} - \mathbf{y} \| = \| \mathrm{A}(\mathrm{QA})^{+}\mathrm{Q}\mathbf{y} - \mathbf{y} \| \\
&\stackrel{(a)}{=} \| \mathrm{U}_{\mathrm{A}}(\mathrm{QU}_{\mathrm{A}})^{+}\mathrm{Q}\mathbf{y} - \mathbf{y} \| \\
&\stackrel{(b)}{=} \| \mathrm{U}_{\mathrm{A}}(\mathrm{QU}_{\mathrm{A}})^{+}\mathrm{Q}(\mathrm{U}_{\mathrm{A}}\mathrm{U}_{\mathrm{A}}^{\mathrm{T}}\mathbf{y} + \boldsymbol{\epsilon}) - \mathrm{U}_{\mathrm{A}}\mathrm{U}_{\mathrm{A}}^{\mathrm{T}}\mathbf{y} - \boldsymbol{\epsilon} \| \\
&\stackrel{(c)}{=} \| \mathrm{U}_{\mathrm{A}}(\mathrm{QU}_{\mathrm{A}})^{+}\mathrm{Q}\boldsymbol{\epsilon} - \boldsymbol{\epsilon} \| \\
&= \| \mathrm{U}_{\mathrm{A}}((\mathrm{QU}_{\mathrm{A}})^{+} - (\mathrm{QU}_{\mathrm{A}})^{\mathrm{T}})\mathrm{Q}\boldsymbol{\epsilon} + \mathrm{U}_{\mathrm{A}}(\mathrm{QU}_{\mathrm{A}})^{\mathrm{T}}\mathrm{Q}\boldsymbol{\epsilon} - \boldsymbol{\epsilon} \| \\
&\stackrel{(d)}{\leq} \| (\mathrm{QU}_{\mathrm{A}})^{+} - (\mathrm{QU}_{\mathrm{A}})^{\mathrm{T}} \| \| \mathrm{Q}\boldsymbol{\epsilon} \| + \| \mathrm{U}_{\mathrm{A}}^{\mathrm{T}}\mathrm{Q}^{\mathrm{T}}\mathrm{Q}\boldsymbol{\epsilon} \| + \| \boldsymbol{\epsilon} \| \\
&\stackrel{(e)}{\leq} \frac{\epsilon}{\sqrt{1-\epsilon}} \| \mathrm{Q}\boldsymbol{\epsilon} \| + \| \mathrm{U}_{\mathrm{A}}^{\mathrm{T}}\mathrm{Q}^{\mathrm{T}}\mathrm{Q}\boldsymbol{\epsilon} \| + \| \boldsymbol{\epsilon} \|.
\end{aligned}$$

(a) follows from Lemma 25; (b) follows from (5); (c) follows Lemma 25, because $\mathrm{QU}_{\mathrm{A}}$ has full rank and so $(\mathrm{QU}_{\mathrm{A}})^{+}\mathrm{QU}_{\mathrm{A}} = \mathrm{I}_d$; (d) follows from the triangle inequality and sub-multiplicativity using $\| \mathrm{U}_{\mathrm{A}} \| = 1$; finally, (e) follows from Lemma 25. We now see the rationale for the complicated sampling probabilities. Since $p_t \geq \epsilon_t^2/\boldsymbol{\epsilon}^{\mathrm{T}}\boldsymbol{\epsilon}$, for $r$ large enough, by Theorem 20, $\| \mathrm{Q}\boldsymbol{\epsilon} \|^2 \leq \| \epsilon \|^2(1+\epsilon)$. Similarly, since $\mathrm{U}_{\mathrm{A}}^{\mathrm{T}}\boldsymbol{\epsilon} = 0$, $\| \mathrm{U}_{\mathrm{A}}^{\mathrm{T}}\mathrm{Q}^{\mathrm{T}}\mathrm{Q}\boldsymbol{\epsilon} \| = \| \mathrm{U}_{\mathrm{A}}^{\mathrm{T}}\boldsymbol{\epsilon} - \mathrm{U}_{\mathrm{A}}^{\mathrm{T}}\mathrm{Q}^{\mathrm{T}}\mathrm{Q}\boldsymbol{\epsilon} \|$; so, we can apply Lemma 17 with $\mathrm{S}_1 = \mathrm{I}_d$, $\mathrm{V} = \boldsymbol{\epsilon}/\| \boldsymbol{\epsilon} \|$ and $\mathrm{S}_2 = \| \boldsymbol{\epsilon} \|$. According to Lemma 17, if $p_t \geq \beta(\mathbf{u}_t^2 + \epsilon_t^2/\boldsymbol{\epsilon}^{\mathrm{T}}\boldsymbol{\epsilon})/(d+1)$, then if $r$ is large enough, $\| \mathrm{U}_{\mathrm{A}}^{\mathrm{T}}\mathrm{Q}^{\mathrm{T}}\mathrm{Q}\boldsymbol{\epsilon} \| \leq \epsilon \| \boldsymbol{\epsilon} \|$. Since these are all probabilistic statements, we need to apply the union bound to ensure that all of them hold. Ultimately, we have:

**Theorem 26.** *For sampling probabilities satisfying (6), and for* $r \geq (8(d+1)/\beta\epsilon^2)\ln\frac{2(d+1)}{\delta}$*, let* $\hat{\mathbf{x}} = (\mathrm{QA})^{+}\mathrm{Q}\mathbf{y}$ *be the approximate regression. Then, with probability at least* $1 - 3\delta$*,*

$$\| \mathrm{A}\hat{\mathbf{x}} - \mathbf{y} \| \leq \left(1 + \epsilon + \epsilon\sqrt{\frac{1+\epsilon}{1-\epsilon}}\right) \| \mathrm{A}\mathbf{x}^* - \mathbf{y} \|,$$

*where* $\mathbf{x}^* = \mathrm{A}^{+}\mathbf{y}$ *is the optimal regression.*

**Remarks.** For the proof of the theorem, we observe that any transformation matrix $Q$ satisfying the following three properties with high probability will do:

$$(i) \| \mathrm{I} - \mathrm{U}^{\mathrm{T}}\mathrm{Q}^{\mathrm{T}}\mathrm{QU} \| \leq \epsilon; \qquad (ii) \| \mathrm{Q}\boldsymbol{\epsilon} \| \leq (1+\epsilon)\| \boldsymbol{\epsilon} \|; \qquad (iii) \| \mathrm{U}^{\mathrm{T}}\mathrm{Q}^{\mathrm{T}}\mathrm{Q}\boldsymbol{\epsilon} \| \leq \epsilon \| \boldsymbol{\epsilon} \|.$$

We will see later that Johnson-Lindenstrauss transforms also satisfy this property, and hence can also be used to perform approximate linear regression with relative error guarantees.

# 7 Approximating the Sampling Probabilities

We have encountered a variety of row sampling probabilities (actually, conditions which the probabilities need to satisfy). Once we can compute probabilities satisfying these conditions, it is relatively straightforward to sample rows according to these probabilities. We now discuss how to efficiently approximate such probabilities.

## 7.1 Matrix Multiplication

The row-norm based sampling is relatively straightforward for the symmetric product. For the asymmetric product, $A^TB$, we need probabilities

$$p_t \geq \beta \frac{\frac{1}{\|A\|^2}\mathbf{a}_t^T\mathbf{a}_t + \frac{1}{\|B\|^2}\mathbf{b}_t^T\mathbf{b}_t}{\rho_A + \rho_B}.$$

To get these probabilities, we need $\|A\|$ and $\|B\|$; since we can compute the exact product in $O(md_1d_2)$, a practically useful algorithm would need to estimate $\|A\|$ and $\|B\|$ efficiently. Suppose we had estimates $\lambda_A, \lambda_B$ which satisfy:

$$(1-\epsilon)\|A\|^2 \leq \lambda_A^2 \leq (1+\epsilon)\|A\|^2; \qquad (1-\epsilon)\|B\|^2 \leq \lambda_B^2 \leq (1+\epsilon)\|B\|^2.$$

We can construct probabilities satisfying the desired property with $\beta = (1-\epsilon)/(1+\epsilon)$.

$$\begin{aligned} p_t &= \frac{\frac{1}{\lambda_A^2}\mathbf{a}_t^T\mathbf{a}_t + \frac{1}{\lambda_B^2}\mathbf{b}_t^T\mathbf{b}_t}{\|A\|_F^2/\lambda_A^2 + \|B\|_F^2/\lambda_B^2} \\ &\geq \frac{\frac{1}{(1+\epsilon)\|A\|^2}\mathbf{a}_t^T\mathbf{a}_t + \frac{1}{(1+\epsilon)\|A\|^2}\mathbf{b}_t^T\mathbf{b}_t}{\|A\|_F^2/(1-\epsilon)\|A\|^2 + \|B\|_F^2/(1-\epsilon)\|A\|^2} \\ &= \left(\frac{1-\epsilon}{1+\epsilon}\right)\frac{\frac{1}{\|A\|^2}\mathbf{a}_t^T\mathbf{a}_t + \frac{1}{\|B\|^2}\mathbf{b}_t^T\mathbf{b}_t}{\rho_A + \rho_B}. \end{aligned}$$

One practical way to obtain $\|A\|^2$ is using the power iteration. Given an arbitrary unit vector $\mathbf{x}_0$, for $n \geq 1$, let $\mathbf{x}_n = A^TA\mathbf{x}_{n-1}/\|A^TA\mathbf{x}_{n-1}\|$. Note that multiplying by $A^TA$ can be done in $O(2md_1)$ operations. By definition, $\|A^TA\mathbf{x}_n\| \leq \|A\|^2$. We now get a lower bound. Let $\mathbf{x}_0$ be a random isotropic vector constructed using $d_1$ independent standard Normal variates $z_1, \ldots, z_{d_1}$; so $\mathbf{x}_0^T = [z_1, \ldots, z_{d_1}]/\sqrt{z_1^2 + \cdots + z_{d_1}^2}$. Let $\lambda_n^2 = A^TA\mathbf{x}_n$ be an estimate for $\|A\|^2$ after $n$ power iterations.

**Theorem 27.** *For some constant $c \leq (\frac{2}{\pi}+2)^3$, with probability at least $1-\delta$,*

$$\lambda_n^2 \geq \frac{\|A\|^2}{\sqrt{4 + \frac{cd_1}{\delta^3}\cdot 2^{-n}}}.$$

**Remarks.** $n \geq \log\frac{d_1}{\delta}$ gives the desired constant factor approximation. Thus in $O(md_1 \ln\frac{d_1}{\delta})$ time, we obtain a sufficiently good estimate for $\|A\|$ (and similarly for $\|B\|$).

*Proof.* Assume that $\mathbf{x}_0 = \sum_{i=1}^{d_1}\alpha_i\mathbf{v}_i$, where $\mathbf{v}_i$ are the eigenvectors of $A^TA$ with corresponding eigenvalues $\sigma_1^2 \geq \cdots \geq \sigma_{d_1}^2$. Note $\|A\|^2 = \sigma_1^2$. If $\sigma_{d_1}^2 \geq \sigma_1^2/2$, then it trivially follows that $\|A^TA\mathbf{x}_n\| \geq \sigma_1^2/2$ for any $n$, so assume that $\sigma_{d_1}^2 < \sigma_1^2/2$. We can thus partition the singular values into those at least $\sigma_1^2/2$ and those which are smaller; the latter set is non-empty. So assume for some $k < d_1$, $\sigma_k^2 \geq \sigma_1^2/2$ and $\sigma_{k+1}^2 < \sigma_1^2/2$. We therefore have:

$$\begin{aligned}
\lambda_n^4 &= \| \mathrm{A}^{\mathrm{T}}\mathrm{A}\mathbf{x}_n \|^2 = \frac{\sum_{i=1}^{d_1} \alpha_i^2 \sigma_i^{4(n+1)}}{\sum_{i=1}^{d_1} \alpha_i^2 \sigma_i^{4n}} \\
&\geq \frac{\sum_{i=1}^{k} \alpha_i^2 \sigma_i^{4(n+1)}}{\sum_{i=1}^{d_1} \alpha_i^2 \sigma_i^{4n}} = \frac{\sum_{i=1}^{k} \alpha_i^2 \sigma_i^{4(n+1)}}{\sum_{i=1}^{k} \alpha_i^2 \sigma_i^{4n} + \sum_{i=k+1}^{d_1} \alpha_i^2 \sigma_i^{4n}}, \\
&= \sigma_1^4 \frac{\sum_{i=1}^{k} \alpha_i^2 (\sigma_i/\sigma_1)^{4(n+1)}}{\sum_{i=1}^{k} \alpha_i^2 (\sigma_i/\sigma_1)^{4n} + \sum_{i=k+1}^{d_1} \alpha_i^2 (\sigma_i/\sigma_1)^{4n}}, \\
&\overset{(a)}{\geq} \sigma_1^4 \frac{\sum_{i=1}^{k} \alpha_i^2 (\sigma_i/\sigma_1)^{4(n+1)}}{4\sum_{i=1}^{k} \alpha_i^2 (\sigma_i/\sigma_1)^{4(n+1)} + 2^{-n}}, \\
&= \frac{\sigma_1^4}{4 + 2^{-n}/\sum_{i=1}^{k} \alpha_i^2 (\sigma_i/\sigma_1)^{4(n+1)}}, \\
&\overset{(b)}{\geq} \frac{\sigma_1^4}{4 + 2^{-n}/\alpha_1^2}.
\end{aligned}$$

(a) follows because for $i \geq k+1$, $\sigma_i^2 < \sigma_1^2/2$; for $i <= k$, $\sigma_1^2/\sigma_i^2 \leq 4$; and $\sum_{i \geq k+1} \alpha_i^2 \leq \sum_{i \geq 1} \alpha_i^2 = 1$. (b) follows because $\sum_{i=1}^{k} \alpha_i^2 (\sigma_i/\sigma_1)^{4(n+1)} \geq \alpha_1^2$. The theorem will now follow if we show that with probability at least $1 - c\delta^{1/3}$, $\alpha_1^2 \geq \delta/d$. It is clear that $\mathbb{E}[\alpha_1^2] = 1/d$ from isotropy. Without loss of generality, assume $\mathbf{v}_1$ is aligned with the $z_1$ axis. So $\alpha_1^2 = z_1^2/\sum_i z_i^2$ ($z_1, \ldots, z_d$ are independent standard normals). For $\delta < 1$, we estimate $\mathbb{P}[\alpha_1^2 \geq \delta/d]$ as follows:

$$\begin{aligned}
\mathbb{P}\left[\alpha_1^2 \geq \frac{\delta}{d}\right] &= \mathbb{P}\left[\frac{z_1^2}{\sum_i z_i^2} \geq \frac{\delta}{d}\right] = \mathbb{P}\left[z_1^2 \geq \frac{\delta}{d}\sum_i z_i^2\right] = \mathbb{P}\left[z_1^2 \geq \frac{\delta}{d-\delta}\sum_{i\geq 2} z_i^2\right] \\
&\geq \mathbb{P}\left[z_1^2 \geq \frac{\delta}{d-1}\sum_{i\geq 2} z_i^2\right] \\
&\overset{(a)}{=} \mathbb{P}\left[\chi_1^2 \geq \frac{\delta}{d-1}\chi_{d-1}^2\right], \\
&\overset{(b)}{\geq} \mathbb{P}\left[\chi_1^2 \geq \delta + \delta^{2/3}\right] \cdot \mathbb{P}\left[\frac{\delta}{d-1}\chi_{d-1}^2 \leq \delta + \delta^{2/3}\right].
\end{aligned}$$

In (a) we compute the probability that a $\chi_1^2$ random variable exceeds a multiple of an independent $\chi_{d-1}^2$ random variable, which follows from the definition of the $\chi^2$ distribution as a sum of squares of independent standard normals. (b) follows from independence and because one particular realization of the event in (a) is when $\chi_1^2 \geq \delta + \delta^{2/3}$ and $\chi_{d-1}^2 \leq \delta + \delta^{2/3}$. Since $\mathbb{E}[\chi_{d-1}^2/(d-1)] = 1$, and $Var[\chi_{d-1}^2/(d-1)] = 2/(d-1)$, by Chebyshev's inequality,

$$\mathbb{P}\left[\frac{\delta}{d-1}\chi_{d-1}^2 \leq \delta + \delta^{2/3}\right] \geq 1 - \frac{2\delta^{1/3}}{d-1}.$$

From the definition of the $\chi_1^2$ distribution, we can bound $\mathbb{P}[\chi_1^2 \leq \delta + \delta^{2/3}]$,

$$\mathbb{P}[\chi_1^2 \leq \delta + \delta^{2/3}] = \frac{1}{2^{1/2}\Gamma(1/2)} \int_0^{\delta+\delta^{2/3}} du\ u^{-1/2} e^{-u/2} \leq \sqrt{\frac{2}{\pi}}(\delta + \delta^{2/3})^{1/2},$$

and so

$$\mathbb{P}\left[\alpha_1^2 \geq \frac{\delta}{d}\right] \geq \left(1-\sqrt{\frac{2}{\pi}}(\delta+\delta^{2/3})^{1/2}\right)\cdot\left(1-\frac{2\delta^{1/3}}{d-1}\right) \geq 1-\left(\frac{2}{\pi}+2\right)\delta^{1/3}.$$

■

We end this section with an alternate sampling based approach to estimate the spectral norm, which can be combined with the power iteration. The basic idea is to do a pre-sampling of the rows of A according to the row norms to construct $\tilde{\mathrm{A}}$. We know that if $r \geq (4\rho_A/\beta\epsilon^2)\ln\frac{2d_1}{\delta}$, then

$$\|\,\tilde{\mathrm{A}}^{\mathrm{T}}\tilde{\mathrm{A}} - \mathrm{A}^{\mathrm{T}}\mathrm{A}\,\| \leq \epsilon\|\,\mathrm{A}\,\|^2.$$

It follows that we have a $\epsilon$-approximation to the spectral norm from

$$\begin{aligned}\|\,\tilde{\mathrm{A}}^{\mathrm{T}}\tilde{\mathrm{A}}\,\| &= \|\,\tilde{\mathrm{A}}^{\mathrm{T}}\tilde{\mathrm{A}} - \mathrm{A}^{\mathrm{T}}\mathrm{A} + \mathrm{A}^{\mathrm{T}}\mathrm{A}\,\| \leq (1+\epsilon)\|\,\mathrm{A}\,\|^2;\\ \|\,\mathrm{A}^{\mathrm{T}}\mathrm{A}\,\| &= \|\,\mathrm{A}^{\mathrm{T}}\mathrm{A} - \tilde{\mathrm{A}}^{\mathrm{T}}\tilde{\mathrm{A}} + \tilde{\mathrm{A}}^{\mathrm{T}}\tilde{\mathrm{A}}\,\| \leq \epsilon\|\,\mathrm{A}\,\|^2 + \|\,\tilde{\mathrm{A}}^{\mathrm{T}}\tilde{\mathrm{A}}\,\|.\end{aligned}$$

Thus, $(1-\epsilon)\|\,\mathrm{A}\,\|^2 \leq \|\,\tilde{\mathrm{A}}^{\mathrm{T}}\tilde{\mathrm{A}}\,\| \leq (1+\epsilon)\|\,\mathrm{A}\,\|^2$. Along this route, one must first sample $r$ rows, and then approximate the spectral norm of the resulting $\tilde{\mathrm{A}}$. We may use the power iteration above to get a constant factor approximation (which is all we need), or we may compute exaactly in $O(rd_1^2)$.

## 7.2 Sparse Matrix Representation

The sampling probabilities $p_t$ are required to satisfy

$$p_t \geq \frac{\beta}{d}\mathbf{u}_t^2.$$

At first sight, it appears that we would need to perform an SVD to merely obtain these sampling probabilities. However, since we only need to approximate these sampling probabilities, we can leverage some recent results from the use of random projections for matrix algorithms Sarlos (2006). In a very active current stream of research, starting with Johnson-Lindenstrauss' original result on embeddings via random projections (Johnson and Lindenstrauss, 1984; Achlioptas, 2003), all the algorithms described here can be accomplished in comparable space and time complexity using fast Johnson-Lindenstrauss transforms, FJLTs, (Ailon and Chazelle, 2006; Sarlos, 2006). In that context, the algorithms achieve their goal by quickly constructing a small number of random linear combinations of the rows. Our focus is to quickly construct a small number of actual rows. Nevertheless, those techniques which quickly construct a small number of linear combinations are usefull for *approximating* the probabilities which can then be used to construct the desired actual rows.

First, some brief background. Let $V = \{\mathbf{v}_1, \ldots, \mathbf{v}_d\}$ where $\mathbf{v}_i \in \mathbb{R}^m$ and $|V| = d$. A matrix $\mathrm{R} \in \mathbb{R}^{r\times m}$ is a Johnson-Lindenstrauss transform (JLT) for $V$ if, for all $\mathbf{x} \in V$,

$$(1-\epsilon)\|\,\mathbf{x}\,\|^2 \leq \|\,\mathrm{R}\mathbf{x}\,\| \leq (1+\epsilon)\|\,\mathbf{x}\,\|^2. \tag{11}$$

We generally assume $m \gg d$. The seminal result of Johnson and Lindenstrauss (1984) is that there exist such transforms with $r = O(\frac{1}{\epsilon^2}\log d)$. Further, it is possible to find such matrices efficiently using randomized constructions. Specifically, if $r = \Omega(\frac{1}{\epsilon^2}\log d \log\frac{1}{\delta})$, then a normalized random matrix (standard normals or random signs (Arriaga and Vempala, 2006)) will yield such a JLT with probability at least $1-\delta$. Ailon and Chazelle (2006) showed that by preconditioning with a randomized Hadamard matrix, a significantly sparser JLT can be constructed, which meant that the matrix multiplications can be computed more efficiently using FFT techniques; we denote this a fast Johnson-Lindenstrauss transform, or FJLT. We will use FLJT to refer to a specific FLJT or, when the context is clear, to denote a probability distribution over $\mathbb{R}^{r\times m}$ which generates an FLJT with high probability. Specifically, it is possible to construct $\mathrm{R} \in \mathbb{R}^{r\times m}$ with $r \geq \frac{c}{\epsilon^2}\log d \log\frac{1}{\delta}$ such that with probability at least $1-\delta$, (11) holds; and, for all $\mathbf{v} \in \mathbb{R}^m$, computing $\mathrm{R}\mathbf{v}$ takes $O(m\log m + r\log^2 d)$ time. This means that for a matrix $\mathrm{A} \in \mathbb{R}^{m\times d}$, we can construct RA in $O(md\log m + rd\log^2 d)$ time.

Let $\mathrm{A} \in \mathbb{R}^{m\times d}$ have SVD $\mathrm{A} = \mathrm{USV}^{\mathrm{T}}$; we also write $\mathrm{U}^{\mathrm{T}} = [\mathbf{u}_1, \ldots, \mathbf{u}_m]$. Let $\mathbf{e}_1, \ldots, \mathbf{e}_m$ be the standard basis vectors in $\mathbb{R}^m$. The columns of U together with the standard basis vectors form a set of $m+d$ vectors. We can project down to $O(\frac{1}{\epsilon^2}\log(d+m))$ dimensions and still preserve the relationships (including angles) between all these vectors Magen (2002). Further, if we project down to $O(\frac{1}{\epsilon^2}d\log\frac{d}{\epsilon})$ dimensions, then RU remains near orthonormal. Summarizing this discussion, together with using Lemmas 6,10 and Corollary 11 in Sarlos (2006), we have the following lemma.

**Lemma 28.** *Let* $\mathrm{R} \in \mathbb{R}^{r\times m}$ *be a FJLT, with* $r \geq \frac{c}{\epsilon^2}(d\log\frac{d}{\epsilon} + \log(d+m))\log\frac{1}{\delta}$. *Then, with probability at least* $1-\delta$, *all the following hold:*

*(i) For* $i = 1, \ldots, m$,

$$\|\,\mathrm{I} - \mathrm{U}^{\mathrm{T}}\mathrm{R}^{\mathrm{T}}\mathrm{RU}\,\| \leq \epsilon; \qquad \|\,\mathrm{R}\mathbf{e}_i\,\| \leq (1+\epsilon); \qquad \|\,\mathbf{e}_i^{\mathrm{T}}\mathrm{UU}^{\mathrm{T}}\mathrm{R}^{\mathrm{T}}\mathrm{R}\mathbf{e}_i - \mathbf{e}_i^{\mathrm{T}}\mathrm{UU}^{\mathrm{T}}\mathbf{e}_i\,\| \leq \epsilon\|\,\mathrm{U}^{\mathrm{T}}\mathbf{e}_i\,\|.$$

*(ii) For* $\mathrm{X} \in \mathbb{R}^{m\times d}$, RX *can be computed in* $O(md\log m + rd\log^2 d)$*; for* $\mathrm{X} \in \mathbb{R}^{d\times r}$, XR *can be computed in* $O(md\log m + rd\log^2 d)$.

Note that the first part of (ii) is from Ailon and Chazelle (2006); the second part also follows exactly the same reasoning in Ailon and Chazelle (2006) using the fact that $\mathrm{R} = \mathrm{PHD}$, where $P$ is sparse, $H$ is Hadamard and D is diagonal. Note that FLJTs can be used to directly perform matrix multiplication, get sparse matrix representations and do linear regression (Sarlos, 2006). All these algorithms would be in terms of a small number of linear combinations of the rows, not the actual rows themselves. The first part of (i) in Lemma 28 is all that one needs for Lemma 25 to hold.

**Lemma 29.** *Under the assumptions of Lemma 28, with probability at least* $1-\delta$, *all of the following hold:*

$$\mathrm{rank}(\mathrm{RA}) = \mathrm{rank}(\mathrm{RU_A}) = \mathrm{rank}(\mathrm{U_A}) = \mathrm{rank}(\mathrm{A}); \tag{12}$$
$$\|\,\mathrm{S_{RU_A}} - \mathrm{S_{RU_A}^{-1}}\,\| \leq \epsilon/\sqrt{1-\epsilon}; \tag{13}$$
$$\|\,(\mathrm{RU_A})^{+} - (\mathrm{RU_A})^{\mathrm{T}}\,\| \leq \epsilon/\sqrt{1-\epsilon}; \tag{14}$$
$$(\mathrm{RA})^{+} = \mathrm{V_A S_A^{-1}}(\mathrm{RU_A})^{+}. \tag{15}$$

### 7.2.1 Constructing The Probabilities

We need to estimate $\mathbf{u}_t^2$. Observe that

$$\mathbf{u}_t^2 = \mathbf{e}_t^{\mathrm{T}}\mathrm{UU}^{\mathrm{T}}\mathbf{e}_t = \mathbf{e}_t^{\mathrm{T}}\mathrm{AA}^{+}\mathbf{e}_t.$$

The costly part here is the computation of $\mathrm{A}^{+}$, and this is where the FJLT comes in. Let R be a FJLT as constructed in Lemma 28. Then, the previous discussion suggests that the projection by R should preserve the angles between $\mathbf{a}_t$ and $\mathrm{A}^{+}\mathbf{e}_t$. Thus, we expect

$$\mathrm{AA}^{+}\mathbf{e}_t \approx \mathrm{A}(\mathrm{RA})^{+}\mathrm{R}\mathbf{e}_t.$$

Hence, we could approximate $\mathbf{u}_t^2$ by $\mathbf{e}_t^{\mathrm{T}}\mathrm{A}(\mathrm{RA})^+\mathrm{R}\mathbf{e}_t$. The only thing we have to be careful about is that this estimate is not necessarily positive, so we will threshold it from below. The algorithm to compute the probabilities is given by

1: Set $\epsilon = \frac{1}{2}\sqrt{\frac{d\log^2 m}{m}}$ in Lemma 28 to obtain R.
2: Compute $\mathrm{X} = (\mathrm{RA})^+\mathrm{R}$.
3: **for** $t = 1\ldots, m$ **do**
4: Compute the estimate $\tilde{w}_t = \mathbf{a}_t^{\mathrm{T}}\mathrm{X}^{(t)}$.
5: Set $w_t = \max\{\epsilon^2, \tilde{w}_t\}$.
6: **end for**
7: $p_t = w_t/\sum_t w_t$ (Normalizing).

For the algorithm described, we obtain the following bound on for the probabilities $p_t$.

**Theorem 30.** *For* $m \geq \frac{4}{9}d\log^2 m$,

$$p_t \geq \frac{1}{c\log^2 m}\frac{\mathbf{u}_t^2}{d}.$$

**Remarks.** The requirement on $m$ is simply to ensure that $\epsilon \leq \frac{3}{4}$; it is benign and would be satisfied provided that $m = \omega(d\log^2 d)$. It is typically the case that $m \gg d$. We note that for application in our matrix algorithms, $\beta = O(1/\log^2 m)$; since the row complexity and computational complexity of all the algorithms is $O(1/\beta)$, this will bloat those by a factor of $\log^2 m$. As we will see our algorithm constructs the approximation $p_t$ in $o(md^2)$ under certain restritions on $m$; it is open whether the probabilities can be approximated better that $O(1/\log^2 m)$ in $o(md^2)$ time

**Computational Complexity** By the choice of $\epsilon$, from Lemma 28, we need $r \geq c\frac{m}{\log m}\log\frac{1}{\delta}$ (where we assume $m \gg d$). From Lemma 28, computing RA takes $O(md(\log m + \log d\log\frac{1}{\delta}))$ time to compute, and the pseudo-inverse takes $O(rd^2) = O(\frac{md^2}{\log m}\log\frac{1}{\delta})$; computing $(\mathrm{RA})^+\mathrm{R}$ also takes $O(md(\log m + \log d\log\frac{1}{\delta}))$ time; finally, given X, computing $p_t$ takes $O(d)$ for a total $O(md)$. Thus, we have a total complexity of

$$O\left(md\log m + (md\log d + \tfrac{md^2}{\log m})\log\tfrac{1}{\delta}\right).$$

For suitable $m$, i.e. $\log m = o(d)$, this complexity is $o(md^2)$.

### 7.2.2 Bounding the Accuracy of the Probability Estimates

We now prove Theorem 30; we show that the estimate $p_t$ constructed in the previous section does indeed produce probabilities which estimate $\mathbf{u}_t^2/d$ up to logarithmic factors. We assume that in the construction of R for the algorithm, $r$ is set according to Lemma 28.

**Lemma 31.** *For* $m \geq \frac{4}{9}d\log^2 m$ *and a constant* $c \geq \frac{7}{2}$,

$$\|\,\mathbf{u}_t\,\|^2 - c\epsilon\|\,\mathbf{u}_t\,\| \leq \tilde{w}_t \leq \|\,\mathbf{u}_t\,\|^2 + c\epsilon\|\,\mathbf{u}_t\,\|.$$

*Proof.* For an arbitrary basis vector $\mathbf{e}_t$, define $\mathbf{e}_t^* = \mathrm{AA}^+\mathbf{e}_t$ which is the best approximation to $\mathbf{e}_t$ in the column space of A. Since $\mathrm{AA}^+$ is a projection operator, $\mathbf{e}_t^{\mathrm{T}}\mathbf{e}_t^* = \mathbf{e}_t\mathrm{UU}^{\mathrm{T}}\mathbf{e}_t = \|\,\mathbf{e}_t^*\,\|^2$. Let $\tilde{\mathbf{e}}_t = \mathrm{A}(\mathrm{RA})^{+}\mathrm{R}\mathbf{e}_t$, and consider $\| \mathbf{e}_t^{\mathrm{T}}\tilde{\mathbf{e}}_t - \mathbf{e}_t^{\mathrm{T}}\mathbf{e}_t^{*} \|$.

$$
\begin{aligned}
\| \mathbf{e}_t^{\mathrm{T}}\tilde{\mathbf{e}}_t - \mathbf{e}_t^{\mathrm{T}}\mathbf{e}_t^{*} \| &\overset{(a)}{=} \| \mathbf{e}_t^{\mathrm{T}}\mathrm{U}(\mathrm{RU})^{+}\mathrm{R}\mathbf{e}_t - \mathbf{e}_t^{\mathrm{T}}\mathrm{U}\mathrm{U}^{\mathrm{T}}\mathbf{e}_t \| \\
&= \| \mathbf{e}_t^{\mathrm{T}}\mathrm{U}((\mathrm{RU})^{+} - (\mathrm{RU})^{\mathrm{T}} + (\mathrm{RU})^{\mathrm{T}})\mathrm{R}\mathbf{e}_t - \mathbf{e}_t^{\mathrm{T}}\mathrm{U}\mathrm{U}^{\mathrm{T}}\mathbf{e}_t \| \\
&\le \| \mathbf{e}_t^{\mathrm{T}}\mathrm{U} \| \| (\mathrm{RU})^{+} - (\mathrm{RU})^{\mathrm{T}} \| \| \mathrm{R}\mathbf{e}_t \| + \| \mathbf{e}_t^{\mathrm{T}}\mathrm{U}\mathrm{U}^{\mathrm{T}}\mathrm{R}^{\mathrm{T}}\mathrm{R}\mathbf{e}_t - \mathbf{e}_t^{\mathrm{T}}\mathrm{U}\mathrm{U}^{\mathrm{T}}\mathbf{e}_t \| \\
&\overset{(b)}{\le} \frac{\epsilon(1+\epsilon)}{\sqrt{1-\epsilon}} \| \mathrm{U}^{\mathrm{T}}\mathbf{e}_t \| + \| \mathbf{e}_t^{\mathrm{T}}\mathrm{U}\mathrm{U}^{\mathrm{T}}\mathrm{R}^{\mathrm{T}}\mathrm{R}\mathbf{e}_t - \mathbf{e}_t^{\mathrm{T}}\mathrm{U}\mathrm{U}^{\mathrm{T}}\mathbf{e}_t \| \\
&\overset{(c)}{\le} \frac{\epsilon(1+\epsilon)}{\sqrt{1-\epsilon}} \| \mathrm{U}^{\mathrm{T}}\mathbf{e}_t \| + \epsilon \| \mathrm{U}^{\mathrm{T}}\mathbf{e}_t \| \\
&\overset{(d)}{\le} c\epsilon \| \mathbf{u}_t \|.
\end{aligned}
$$

(a) and (b) follow from Lemma 29; (c) from Lemma 28; and, (d) assumes that $\epsilon \le \frac{3}{4}$ which follows from the definition of $\epsilon$ and the condition assumed on $m$. Notice that $\mathbf{e}_t^{\mathrm{T}}\tilde{\mathbf{e}}_t$ is exactly the estimate $\tilde{\mathbf{w}}_t$, and $\mathbf{e}_t^{\mathrm{T}}\mathbf{e}_t^{*} = \| \mathbf{u}_t \|^2$; the lemma follows. ■

Define $\epsilon' = c\epsilon$. The next corollary is immediate from the lower bound in Lemma 31 and elementary calculus.

**Corollary 32.** $\tilde{w}_t \ge -\frac{1}{4}c^2\epsilon^2$.

It is also useful to invert the bounds in Lemma 31 to get bounds for $\| \mathbf{u}_t \|^2$ in terms of $\tilde{w}_t$.

**Corollary 33.** $\tilde{w}_t + \frac{1}{2}c^2\epsilon^2 - \frac{1}{2}c\epsilon\sqrt{c^2\epsilon^2 + 4\tilde{w}_t} \le \| \mathbf{u}_t \|^2 \le \tilde{w}_t + \frac{1}{2}c^2\epsilon^2 + \frac{1}{2}c\epsilon\sqrt{c^2\epsilon^2 + 4\tilde{w}_t}$

*Proof.* Using the quadratic formula,

$$
\begin{aligned}
\| \mathbf{u}_t \|^2 + c\epsilon \| \mathbf{u}_t \| \ge \tilde{w}_t &\implies (\| \mathbf{u}_t \| + \tfrac{1}{2}c\epsilon + \tfrac{1}{2}\sqrt{c^2\epsilon^2 + 4\tilde{w}_t})(\| \mathbf{u}_t \| + \tfrac{1}{2}c\epsilon - \tfrac{1}{2}\sqrt{c^2\epsilon^2 + 4\tilde{w}_t}) \ge 0; \\
\| \mathbf{u}_t \|^2 - c\epsilon \| \mathbf{u}_t \| \le \tilde{w}_t &\implies (\| \mathbf{u}_t \| - \tfrac{1}{2}c\epsilon + \tfrac{1}{2}\sqrt{c^2\epsilon^2 + 4\tilde{w}_t})(\| \mathbf{u}_t \| - \tfrac{1}{2}c\epsilon - \tfrac{1}{2}\sqrt{c^2\epsilon^2 + 4\tilde{w}_t}) \le 0.
\end{aligned}
$$

In the first inequality, the first term is positive, so the second must also be positive:

$$
\| \mathbf{u}_t \| \ge \frac{1}{2}\sqrt{c^2\epsilon^2 + 4\tilde{w}_t} - \frac{1}{2}c\epsilon.
$$

In the second inequality, if the second term is positive, the first cannot be negative, so it must be that the second term is negative and the first is positive, and so

$$
\begin{aligned}
\| \mathbf{u}_t \| &\ge \tfrac{1}{2}c\epsilon - \tfrac{1}{2}\sqrt{c^2\epsilon^2 + 4\tilde{w}_t}; \\
\| \mathbf{u}_t \| &\le \tfrac{1}{2}c\epsilon + \tfrac{1}{2}\sqrt{c^2\epsilon^2 + 4\tilde{w}_t}.
\end{aligned}
$$

Combining these three inequalities and squaring concludes the proof. ■

The appearence of $\| \mathbf{u}_t \|$ in the bound of Lemma 31 is a serious problem, because it means that $\tilde{w}_t$ is not a relative approximation to $\| \mathbf{u}_t \|^2$, which would have been the ideal situation. This poses a serious problem if we are to normalize the $\tilde{w}_t$ by $\sum_t w_t$ to obtain probabilities. If there are too many small $\mathbf{u}_t$'s which appear large due to the error from the projection, then when we normalize, this inflates the importance of the small $\mathbf{u}_t$'s. This is why in the algorithm, we thresholded the $\tilde{w}_t$'s to obtain the $w_t$'s. Since we will be normalizing by the sum, we will need bounds on the sums.

**Lemma 34.** *For $m \geq \frac{4}{9} d \log^2 m$ and a constant $c \leq 1 + \sqrt{7}$,*

$$(1 - c\epsilon)d \leq \sum_{t=1}^{m} \tilde{w}_t \leq (1 + c\epsilon)d.$$

*Proof.* We use the fact that $\sum_{t=1}^{m} \tilde{w}_t = \sum_{t=1}^{m} \mathbf{e}_t^{\mathrm{T}} \mathrm{U}(\mathrm{RU})^{+}\mathrm{R}\mathbf{e}_t = \mathrm{trace}(\mathrm{U}(\mathrm{RU})^{+}\mathrm{R})$:

$$\begin{aligned}
|\mathrm{trace}(\mathrm{U}(\mathrm{RU})^{+}\mathrm{R}) - d| &= |\mathrm{trace}(\mathrm{U}((\mathrm{RU})^{+} - (\mathrm{RU})^{\mathrm{T}})\mathrm{R}) + \mathrm{trace}(\mathrm{U}(\mathrm{RU})^{\mathrm{T}}\mathrm{R}) - d| \\
&\overset{(a)}{=} |\mathrm{trace}(((\mathrm{RU})^{+} - (\mathrm{RU})^{\mathrm{T}})\mathrm{RU}) + \mathrm{trace}(\mathrm{U}^{\mathrm{T}}\mathrm{R}^{\mathrm{T}}\mathrm{RU}) - d| \\
&\overset{(b)}{\leq} \sqrt{d} \|\, ((\mathrm{RU})^{+} - (\mathrm{RU})^{\mathrm{T}})\mathrm{RU} \,\|_F + |\mathrm{trace}(\mathrm{U}^{\mathrm{T}}\mathrm{R}^{\mathrm{T}}\mathrm{RU}) - d| \\
&\overset{(c)}{\leq} \sqrt{d} \|\, (\mathrm{RU})^{+} - (\mathrm{RU})^{\mathrm{T}} \,\| \|\, \mathrm{RU} \,\|_F + |\mathrm{trace}(\mathrm{U}^{\mathrm{T}}\mathrm{R}^{\mathrm{T}}\mathrm{RU}) - d|.
\end{aligned}$$

(a) is from the cyclic property of the trace. (b) follows from the facts that $\|\, \mathrm{X} \,\|_F^2 = \mathrm{trace}\mathrm{X}^{\mathrm{T}}\mathrm{X}$, and the bound on the trace for rank $d$ matrices obtained using the Cauchy-Schwarz inequality:

$$\mathrm{trace}(\mathrm{X}) = \sum_i \sigma_i(\mathrm{X}) \leq \sum_i |\sigma_i(\mathrm{X})| \leq \sqrt{d} \sqrt{\sum_i \sigma_i^2(\mathrm{X})} = \sqrt{d} \|\, \mathrm{X} \,\|_F.$$

Finally, (c) follows because $\|\, \mathrm{XY} \,\|_F \leq \|\, \mathrm{X} \,\| \|\, \mathrm{Y} \,\|_F$. Note that $\|\, \mathrm{RU} \,\|_F^2 = \mathrm{trace}(\mathrm{U}^{\mathrm{T}}\mathrm{R}^{\mathrm{T}}\mathrm{RU})$. By Lemma 28, $|\sigma_i^2(\mathrm{U}^{\mathrm{T}}\mathrm{R}^{\mathrm{T}}\mathrm{RU}) - 1| \leq \epsilon$; hence $(1 - \epsilon)d \leq \sum_i \sigma_i^2(\mathrm{U}^{\mathrm{T}}\mathrm{R}^{\mathrm{T}}\mathrm{RU}) \leq (1 + \epsilon)d$. Thus, using Lemma 29,

$$|\mathrm{trace}(\mathrm{U}(\mathrm{RU})^{+}\mathrm{R}) - d| \leq d\epsilon \sqrt{\frac{1 + \epsilon}{1 - \epsilon}} + d\epsilon.$$

To complete the proof, note that the condition assumed on $m$ makes $\epsilon \leq \frac{3}{4}$. ∎

We are now ready to complete the proof of Theorem 30. We first recap the definition of $w_t$:

$$w_t = \begin{cases} \epsilon^2 & \tilde{w}_t \leq \epsilon^2; \\ \tilde{w}_t & \tilde{w}_t > \epsilon^2. \end{cases}$$

Since $\tilde{w}_t \geq -\frac{1}{4} c\epsilon^2$, it follows that $w_t \leq \tilde{w}_t + \epsilon^2(1 + \frac{c}{4})$. Thus, using the definition of $\epsilon$,

$$\begin{aligned}
\sum_{t=1}^{m} w_t &\leq (1 + c'\epsilon)d + m\epsilon^2(1 + \tfrac{c}{4}) \\
&= d\left((1 + c'\epsilon) + \tfrac{\log^2 m}{4}(1 + \tfrac{c}{4})\right) \\
&\leq cd \log^2 m
\end{aligned}$$

If $\tilde{w}_t \leq \epsilon^2$, then by Corollary 33 (replacing $\tilde{w}_t$ with $\epsilon^2$ in the upper bound),

$$\begin{aligned}
\|\, \mathbf{u}_t^2 \,\| &\leq \epsilon^2 + \frac{1}{2} c^2 \epsilon^2 (1 + \sqrt{1 + 4/c^2}) \\
&\leq c\epsilon^2 = cw_t.
\end{aligned}$$

If $\tilde{w}_t > \epsilon^2$, then, again, by Corollary 33 (replacing $\epsilon^2$ with $\tilde{w}_t$ in the upper bound),

$$\begin{aligned}
\|\, \mathbf{u}_t^2 \,\| &\leq \tilde{w}_t + \frac{1}{2} c^2 \tilde{w}_t (1 + \sqrt{1 + 4/c^2}) \\
&\leq c\tilde{w}_t = cw_t.
\end{aligned}$$

In both cases, $w_t \geq \frac{1}{c}\mathbf{u}_t^2$, which concludes the proof, because

$$p_t = \frac{w_t}{\sum_{i=1}^m w_t} \geq \frac{1}{c \log^2 m} \frac{\mathbf{u}_t^2}{d}.$$

### 7.3 $\ell_2$ Regression

The probabilities depend on $\mathbf{u}_t^2$ (the leverage scores) and $\epsilon_t$ (the components of the residual error). Though the previous ideas would apply to constructing approximations to the leverage scores, it is not clear how one could get near relative error approximations to the components of the residual error. Indeed, sampling algorithms for $\ell_2$ regression may have non-zero values for estimated residuals where the actual residual is zero. We would certainly need that when the actual residual converges to zero in some components, then so does the estimated residual in those components. Though we can get a $1+\epsilon$ approximation to the sum of squared residuals, this appears to be of not much help to get the squared residuals themselves. An algorithm to construct a good approximation to the actual residuals would then mean that an efficient row-sampling algorithm for the $\ell_2$ regression can also be obtained. As of yet, ithas been elusive.